\newcommand{\CLASSINPUTtoptextmargin}{0.78in}
\documentclass[journal, draftcls, onecolumn]{IEEEtran}
\IEEEoverridecommandlockouts
\usepackage{cite}
\usepackage{amsmath,amssymb,amsfonts}
\usepackage{mathrsfs}
\usepackage{algorithmic}
\usepackage{graphicx}
\usepackage{textcomp}
\usepackage{xcolor}

\usepackage{algorithm}
\usepackage{graphicx}  
\usepackage{float}  
\usepackage{subfigure}  
\usepackage{bm}

\newcommand{\clrtx}[1]{\textcolor{blue}{#1}}

\newcommand{\reffig}[1]{Fig. \ref{#1}}
\newcommand{\refeqn}[1]{(\ref{#1})}

\newtheorem{conjecture}{Conjecture}

\def\BibTeX{{\rm B\kern-.05em{\sc i\kern-.025em b}\kern-.08em
    T\kern-.1667em\lower.7ex\hbox{E}\kern-.125emX}}
\usepackage{indentfirst}
\usepackage{upgreek}
\usepackage{etoolbox}
\makeatletter
\patchcmd{\@makecaption}
{\scshape}
{}
{}
{}
\makeatother
\begin{document}

\title{Levenberg-Marquardt Method Based Cooperative Source Localization in SIMO Molecular Communication via Diffusion Systems
}
\author{ Yuqi Miao, Wence Zhang*, Xu Bao
	\thanks{ Y. Miao, W. Zhang, and X. Bao are with the School of Computer and Communications Engineering, Jiangsu University, Zhenjiang, China (e-mail: {2221808003}@stmail.ujs.edu.cn; \{wencezhang,xbao\}@ujs.edu.cn). Part of this work was presented in 2019 IEEE International Conference on Communication Technology.}
}

\maketitle

\begin{abstract}
Molecular communication underpins nano-scale communications in nanotechnology. The combination of multi-nanomachines to form nano-networks is one of the main enabling methods. Due to the importance of source localization in establishing nano-networks, this paper proposes a cooperative source localization method for Molecular Communication via Diffusion (MCvD) systems using multiple spherical absorption receivers. Since there is no exact mathematical expression of the channel impulse response for multiple absorbing receivers, we adopt an empirical expression and use Levenberg-Marquardt method to estimate the distance of the transmitter to each receiver, based on which the location of the transmitter is obtained using an iterative scheme where the initial point is obtained using a multi-point localization method. Particle based simulation is carried out to evaluate the performance of the proposed method. Simulation results show that the proposed method can accurately estimate the location of transmitter in short to medium communication ranges.
\end{abstract}

\begin{IEEEkeywords}
Localization, distance estimation, molecular communication via diffusion (MCvD), multiple receivers, absorbing spherical receivers.
\end{IEEEkeywords}

\section{Introduction}
The mutual communication between nano-scale devices will make it possible for nanotechnology to be applied in the fields like disease sensing, drug delivery, environmental monitoring and etc \cite{Nakano2012}. Due to the limited size and capabilities of one single nanomachine, a large number of nanomachines that cooperate to form a nanonetwork are attractive for completing complex tasks \cite{Huang2019}. Inspired by nature, molecular communication that uses molecular coding, transmission, and reception of information has become one of the enabling technologies for nanonetworks due to its biocompatibility and energy efficiency\cite{Korte2017,Murat2019}. There are mainly three types of molecular transmission methods, i.e., walkway-based, flow-based and diffusion-based\cite{Farsad2017}, among which molecular communication via diffusion (MCvD)  is the simplest mechanism, where the transmitter nanomachine (TN) encodes and transmits to the receiver nanomachine (RN) using molecules that diffuse freely in the environment through Brownian motion and no external energy is required\cite{Ali2018}. However, the delay and uncertainty of diffusion restrict the communication distance between nanomachines. To increase the communication range,  flow based molecular communication provides an excellent solution\cite{Noel2014}, because molecules rely on not only diffusion, but also fluid movement in the medium which increases the efficiency of molecular motion, e.g., drug delivery in blood vessels.

TN's location is a key factor in MCvD systems. With predictable location of the TN in nanonetworks, the system parameters such as the number of released molecules and the time of arrival can be optimized  accordingly to reduce unnecessary transmission costs and interference (e.g., excessive molecules will generate strong interference). For example, the RN could adjust receiving direction\cite{Felicetti2019} and receiving delay\cite{Akdeniz2018} according to TN's location, so as to reduce interference and achieve optimal reception performance. At the same time, it is also conducive to the layout and location adjustment of the RNs per unit volume \cite{Aijaz2015}. Moreover, in application areas such as targeted drug delivery system, knowing the current location and destination of nanomachine-bound drug is important for its routing \cite{Atakan2012}. In terms of environmental monitoring, it is helpful to determine the location of the pollution source, such as determining the leak point of the subsea oil pipelines \cite{Jamali2019}.

In recent years, research on TN-RN distance estimation methods in MCvD has attracted significant  attention \cite{Huang2013,Luo2018,Moore2012,Wang22015,Wang2015,Lin2018,Guo2016,kose2020}. In \cite{Moore2012}, the authors proposed to use Single Spike Feedback Signals to estimate the signal attenuation strategy and a round trip time (the time for signal transmission of TN-RN-TN), which required complex duplex nanomachines.
In \cite{Huang2013}, one-way strategy was used to estimate the distance which used the time of peak concentration at the RN. However, measuring the precise peak concentration time was no-easy task due to the uncertainty of random molecular motion.
The authors of \cite{Luo2018} improved the work in \cite{Moore2012} by using two kinds of molecules. However, the performance is limited by the intra-reaction between different molecules. The TN-RN distance estimation in three-dimensional (3D) systems has drawn great research interest. In \cite{Wang22015}, the authors used the time of peak concentration and the summation of molecular concentration to estimate TN-RN distance.
In \cite{Wang2015}, TN-RN distance is estimated by periodically calculating the cumulative number of molecules captured by the reciever. The inter-symbol interference will affect the performance of the proposed scheme.
Distance estimation based on maximum likelihood estimation with Newton–Raphson method was proposed in \cite{Lin2018}, which achieves high-accuracy, but had high storage and computational requirements.

Another kind of source location is related to eavesdropper localization, which uses received signals generated by the RN to estimate the eavesdropper's location\cite{Guo2016,kose2020}. In \cite{Guo2016}, the authors considered that when there was only one eavesdropper in a one-dimensional environment, the location of the eavesdropper is determined according to the change in the number of molecules received by the RN. This scheme has been improved in \cite{kose2020}, where the authors built a deep neural network to detect eavesdroppers near the TN and the RN by comparing the difference in the number of molecules received above and below the RN.

 The above research schemes are all used for distance estimation with only one transmitter and one receiver in the environment. Since the channel impulse response in this scenario can be solved by Fick's second law of diffusion (recent results can be found in \cite{Yilmaz2014,Deng2015TMBMC,Genc2018NCN} and references therein), the TN-RN distance can be directly calculated using the time of peak concentration or the cumulative number of molecules captured. However, when there are multiple absorbing receivers in the environment, the explicit expression of the channel impulse response is very difficult to obtain, and there are few theoretical results except for \cite{Kwak2020TCOMM}, where an analytical expression is provided for scenarios with two absorbing receivers. However, TN localization can not be done with only two receivers and the problem remains unsolved.

In this work, we consider TN localization problem in single-input-multiple-output (SIMO) MCvD systems and propose a novel  method to estimate the TN-RN distance aiming to locate the TN. To the best of the authors' knowledge, this work is the first attempt for source localization in 3D SIMO-MCvD systems.

The main contributions of this work are summarized as follows:
\begin{itemize}
	\item We propose a general model for TN localization in SIMO-MCvD systems with consideration of both the RN topology and diffusion environment factors.	
	\item A novel cooperative source localization (CSL) scheme is proposed. We use an empirical expression of the cumulative number of received molecules with an introduced auxiliary parameter, from which the estimated distance between TN and each RN is obtained by Levenberg-Marquardt method. The estimated TN location is achieved by solving an optimization problem using steepest descent method with initial value obtained by multi-point localization.
	\item Extensive particle-based simulations verify the effectiveness of the proposed method in the short to medium communication range. Meanwhile, the influence of different factors on the localization accuracy is also analyzed, i.e., RN's radius, the number of transmitted molecules and the diffusion coefficient. The proposed method can also be applied to scenarios with flow, and better estimation accuracy is obtained under certain conditions. In addition, the computational complexity of the proposed CSL is analyzed and reduced by taking advantage of the maximum sample interval.
    \item The optimal RN topology is discussed. It is conjectured that when TN is uniformly distributed in the environment, a symmetric and distributed topology is preferred.
\end{itemize}

The rest of this paper is organized as follows. In Section II, the SIMO-MCvD system model is introduced and TN localization problem is formulated. The proposed CSL method is explained in Section III. Numerical results are provided in Section IV and conclusions are drawn in Section V.

\emph{Notation:} Vectors and matrices are shown in bold font; $\nabla$ is the gradient operator and  $\nabla^2$ is the Laplace operator; $\left\| \cdot \right\|_2$ denotes L2 norm; $\mathcal N (\mu, \sigma^{2})$ denotes the normal distribution with the mean $\mu$ and a variance of $\sigma^{2}$; $(\cdot)^{\text T}$ denotes the transpose of a vector or a matrix.

\section{Problem Formulation}

\begin{figure}[t]
	\centering
	\includegraphics[width=1.0\linewidth]{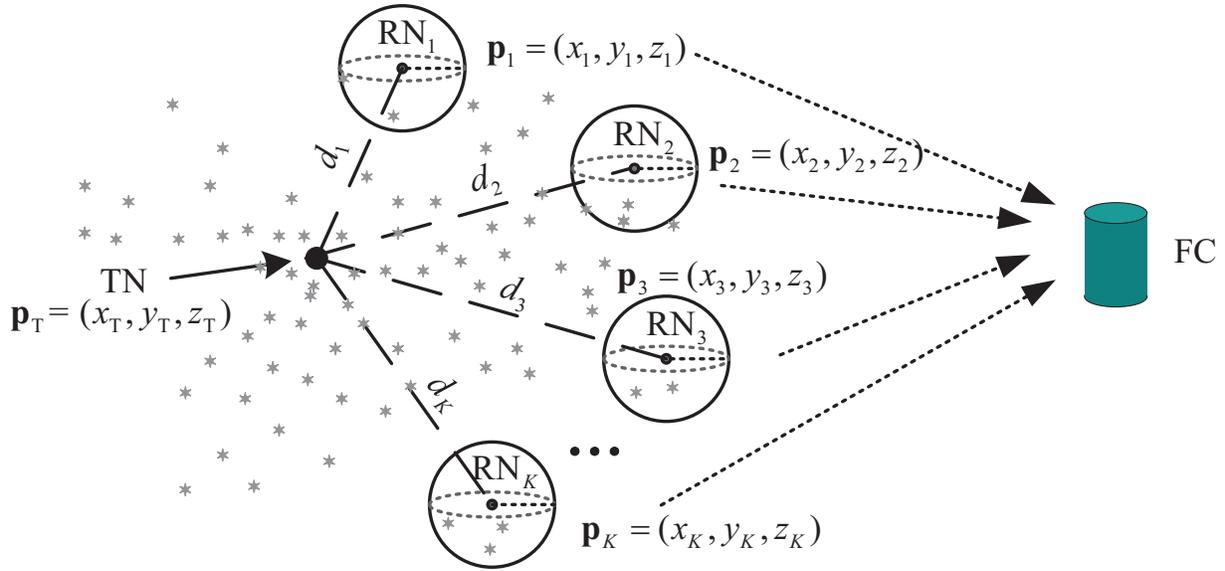}
	\caption{MCvD system model with one point TN and $K$ absorbing spherical RNs.}\label{fig:systemmodel}
\end{figure}

As shown in Fig.\ref{fig:systemmodel}, the SIMO-MCvD system consists of one TN and $K$ RNs in an unbounded 3D environment. The TN is a point source transmitter that can release molecules at a certain moment. It is located at ${\bf{p}}_{\text T}=({x_{\text{T}}},{y_{\text{T}}},{z_{\text{T}}})$. The released information molecules spread as Brownian motion in the environment. The RNs are all absorbing spherical receivers, which have receptor molecules  reacting with information molecules completely on the surface. In other words, if a molecule hits RN, it will be completely absorbed and removed from the environment. \clrtx{ It is assumed that RNs can count the number of received molecules and report to the fusion center (FC) using predefined types of molecules that are different from TN's \cite{Fang2019TCOM}. We also assume that the links between the RNs and the FC are perfect. The FC uses received information from RNs to locate the TN.} The location and the radius of the $k$-th RN are denoted as ${{\bf{p}}_k} = ({x_k},{y_k},{z_k})$ and ${r_k}$, respectively. Let $d_k$ be the distance between TN and ${\text{RN}}_k$, i.e., $d_k = {\left\| {{\bf{p}_{\text{T}}} - {\bf{p}}_k} \right\|_2}$. To focus on the dominant factors, the influence of other molecules in the environment is omitted, and thus the information molecules will not react or degrade with other molecules in the environment.

In MCvD systems, molecules propagate in the environment following Fick's law which was proposed by Fick in 1855\cite{philibert2006}. Fick's first law of diffusion describes the migration of a substance from a high concentration region to a low concentration region, which is expressed as
\begin{equation}\label{eqn:fick1}
\begin{split}
J({\bf{p}},t) & =  - D\nabla C({\bf{p}},t),   \\
      \nabla  & = \frac{\partial }{{\partial x}} + \frac{\partial }{{\partial y}} + \frac{\partial }{{\partial z}},
\end{split}
\end{equation}
where $D$ is the diffusion coefficient; $J({\bf{p}},t)$ and $C({\bf{p}},t)$ are the diffusion flux and molecular concentration at time $t$ and at location  ${\bf{p}} = (x,y,z)$, respectively.

Assuming that there is no chemical reaction in the environment and no interference from other molecules, mass conservation principle can be used to obtain Fick’s second law of diffusion as

\begin{equation}\label{eqn:fick2}
\begin{split}
\frac{{\partial C({\bf{p}},t)}}{{\partial t}} &= D{\nabla ^2}C({\bf{p}},t) ,\\
                \nabla ^2 &=\frac{\partial ^2}{\partial {x^2}} + \frac{\partial ^2}{\partial {y^2}} + \frac{\partial ^2}{\partial {z^2}}.
\end{split}
\end{equation}

In the SIMO-MCvD system shown in Fig.\ref{fig:systemmodel} at time $t$ and at location  ${\bf{p}} = (x,y,z)$, the concentration $C({\bf{p}},t)$ is determined by the following partial differential equations

\begin{subequations}\label{eqn:partial}
\begin{align}
\frac{{\partial C({\bf{p}},t)}}{{\partial t}} =& D{\nabla ^2}C({\bf{p}},t),\\
C({\bf{p}}_{\text T}, 0)=& \delta ({\bf{p}},t),\\
C({\bf{p}},t)=& 0, \quad\text{for} \  { \left\| {\bf{p}} - {{\bf{p}}_{\text{T}}} \right\|}_2 \rightarrow \infty \\
D\nabla C({\bf{p}},t) =& wC({\bf{p}},t),  \quad\text{for}  \ {\bf{p}} \in {{\Omega}_k} \\
&{\kern 1pt} {\kern 1pt} {\kern 1pt} \,k = 1, \ldots ,K\nonumber,
\end{align}	
\end{subequations}
where $\Omega _k$ is the surface of $\text{RN}_k$ which is defined as

\begin{equation}
{\Omega _k} \buildrel \Delta \over =\left\lbrace   {\bf{p}} \in\mathbb{R} {^3} \big|  { \left\| {\bf{p}} - {{\bf{p}}_k} \right\|}_2  = {r_k}    \right\rbrace .
\end{equation}
In \eqref{eqn:partial}, (3b) is the initial condition, indicating that the TN releases molecules at $t=0$ and at location ${\bf{p}}_{\text T}$; $\delta ( \cdot )$  is Dirac delta function. (3c) is the first boundary condition of the diffusion equation where unbound environment is considered; the molecular concentration vanishes as the distance is much greater than TN-RN distance. (3d) is the second boundary conditions with $K$ RNs, and $w$ is the reaction rate of receptor molecules on the surface of RNs with information molecules emitting from TN. In the case of full absorption, the reaction rate is close to infinity, and then $C({\bf{p}}-{\bf{p}}_k,t)=0$. When a flow with velocity vector ${\bf v} ({\bf p},t)$ is considered, the flowing advection equation should be satisfied \cite{Jamali2019}, and (3a) is changed to
\begin{equation*}\label{eqn:advection_eq}
\frac{{\partial C({\bf{p}},t)}}{{\partial t}} + \nabla \cdot [ {\bf v} ({\bf p},t) C({\bf p},t)]  = D{\nabla ^2}C({\bf{p}},t).
\end{equation*}

Solving \eqref{eqn:partial} with or without flow  yields the concentration distribution $C({\bf{p}},t)$ at any location and at any time. Then the average hitting rate at receiver $k$ is expressed as
\begin{equation}\label{eqn:hitting}
{f_k}({\bf{p}}_\text{T},t)=\int_{\Omega _k} {w C({\bf{p}},t) \, \text{d} \bf{p}} .
\end{equation}
By integrating \eqref{eqn:hitting}, the cumulative number of molecules hitting $\Omega _k$ until time $t$ is expressed as
\begin{equation}\label{eqn:CDF}
{F_k}({\bf{p}}_\text{T},t) = \int_0^t {Q {f_k}({\bf{p}}_\text{T},\tau) \, \, {\text{d}}} \tau ,
\end{equation}
where $Q$ denotes the number of molecules emitted by TN.

 \begin{figure}[t]
 	\centering
 	\includegraphics[width=0.95\linewidth]{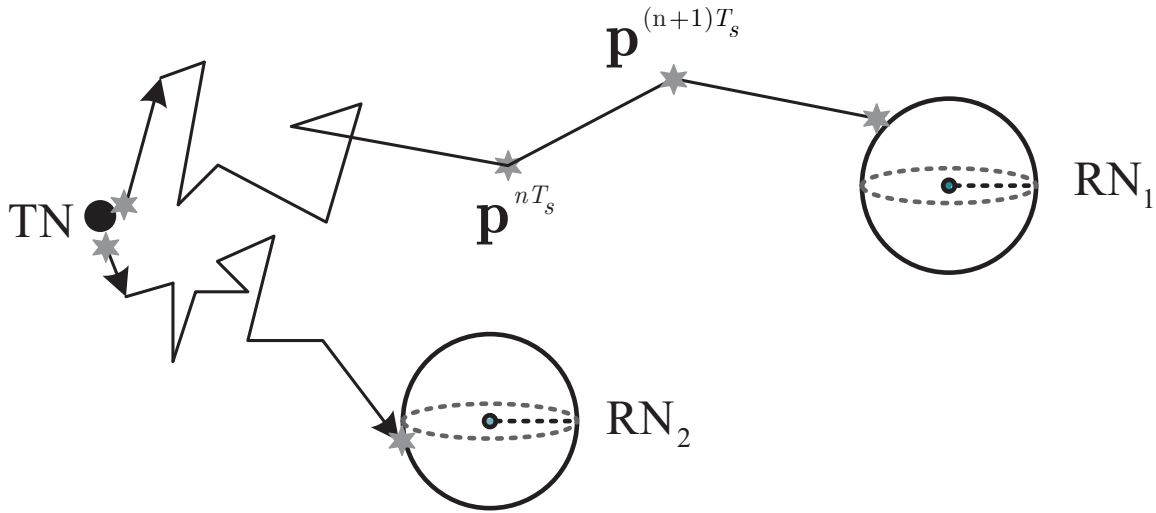}
 	\caption{An illustration of the Brownian motion model.}\label{fig:Brownian}
 \end{figure}

Molecules released by TN spread  as Brownian motion in the environment, as shown in Fig. \ref{fig:Brownian}. We assume that the molecules in the environment move once per time slot to approximate the Brownian motion process, and this time slot is the sampling interval (${T_s}$). Therefore, the location of a molecule at ($n+1$)-th $T_s$ is modeled as
\begin{equation}
{\bf{p}}^{(n+1){T_s}}={\bf{p}}^{n{T_s}}+\Delta\bf{p},
\end{equation}
where $\Delta {\bf p} = {\bf v}({\bf p},t) T_s + \bf n_p$ is the instantaneous displacement of each molecule where $\bf n_p$ follows $\mathcal N (\mu, 2DT_s)$ in each dimension.

Due to the uncertainty of Brownian motion, the actual number of received molecules $\tilde {F}_k$ varies significantly. Therefore, additional noise $I_k $ should be added to the theoretical value in \eqref{eqn:CDF}, which is described as
\begin{equation}
\tilde{F_k}({\bf{p}_\text{T}},n{T_s}) ={F_k}({\bf{p}_\text{T}},n{T_s})+I_k(n{T_s}),
\end{equation}
 where $I_k$ is usually considered to be Gaussian distributed with zero mean and a variance related to the molecular concentration and the RN's volume \cite{Lin2018}.

In MCvD systems, source localization has an important influence on the selection of optimal reception delay \cite{Akdeniz2018}, the threshold for signal detection\cite{Tiwari2017}, etc. Therefore, it is of great importance for the RNs to be aware of source location. The cumulative number of received molecules at each RN is the only information that we could leverage here, based on which, the source localization problem can be modeled as the following least-square problem
\begin{equation}\label{eqn:main}
{\bf{p}}_{\text{T}}^* = \mathop {\arg \min\limits_{{\bf{p}_\text{T}} \in \mathbb{R}^3 }  } \sum\limits_{k = 1}^K {\sum\limits_{n = 1}^{{N_{\rm{s}}}} {\left\| {\tilde{F_k}({\bf{p}_\text{T}},n{T_s}) - {F_k}({\bf{p}_\text{T}},n{T_s})} \right\|_2^2} } ,
\end{equation}
where $N_s$ is the number of samples and ${\bf{p}}_{\text{T}}^*=({x_{\text{T}}^*},{y_{\text{T}}^*},{z_{\text{T}}^*})$ is the estimated location of TN. In \refeqn{eqn:main}, it can be seen that at least three receivers are required for source localization.

 Due to the complex form of the boundary conditions, the partial differential equation has no closed-form solution in \eqref{eqn:partial}, which makes it difficult to obtain the theoretical value of the cumulative number of received molecules in \eqref{eqn:CDF}. This is the main difficulty in solving problem \eqref {eqn:main}.

\section{Cooperative Source Localization (CSL)}
In this section, a novel source localization scheme was proposed for SIMO-MCvD systems.

In order to solve the source localization problem, an empirical expression for $\tilde{F_k}$ with intermediate auxiliary parameters is proposed according to our previous work \cite{Bao2019}. Then,  the parameters of $\tilde{F_k}$ are determined by the Levenberg-Marquardt (LM) algorithm using sampled data of received molecules. By taking advantage of the intermediate variables, we propose to estimate the distance from TN to $\text{RN}_k$ at first, and then use the steepest descent (SD) algorithm with initial value obtained by a multi-point localization scheme to find the location of TN. The above solution is reffered to as the cooperative source localization (CSL) method. \clrtx{Although we consider a point source here, the proposed CSL method can be adapted to other types of TNs by finding proper empirical expressions.}

\subsection{Empirical Expression of ${F_k}$}
It is difficult to derive ${F_k}$ by solving \eqref{eqn:CDF} directly. When $K=1$, i.e., when there is only one RN in the environment, the differential equation can be solved without considering flow. The cumulative number of received molecules by the RN is given by \cite{Yilmaz2014}

\begin{equation}\label{eqn:siso}
F(d,t) = Q\frac{r}{d}{\rm{erfc}}\left( {\frac{{d - r}}{{\sqrt {4Dt} }}} \right),
\end{equation}
where $d$ is the distance between TN and RN and $r$ is the radius of RN. However, there is mutual interplay between each RN with more than one RN in the environment, so \eqref{eqn:siso} is not applicable for SIMO-MCvD systems.

To the best of the authors' knowledge, however, there is no solution to \eqref{eqn:CDF} existing in the literature with or without flow.  In our previous work \cite{Bao2019}, it is projected that ${F_k}$ decreases as the number of RN increases, and is related to many parameters, such as $d_k$, $r_k$, $K$, etc. Since the molecules diffuse in the environment as a ``random walk'' like process, the step length of which is usually modeled by Gaussian distribution. Therefore, it is almost surely $F$ has a form containing ``erfc'' functions. In order to capture the leading factors, we introduce an auxiliary parameter $a_k$ into \eqref{eqn:siso} and formulate the cumulative number of received molecules as

\begin{equation}\label{eqn:fit}
{F_k}_{\text{fit}}({{\bf{p}}}_\text{T},t) = {a_k}Q\frac{r_k}{{{d_k}}}{\rm{erfc}}\left( {\frac{{{d_k} - {r_k}}}{{\sqrt {4Dt} }}} \right).
\end{equation}
For each RN,  $F_k$ can be well approximated by \eqref{eqn:fit} where $a_k$ and $d_k$ are obtained by the Levenberg-Marquardt method. The introduced auxiliary parameter $a_k$ is also able to cope with scenarios with flow, as will be shown in the numerical results.

\subsection{Levenberg-Marquardt Method Based Distance Estimation}
There are many methods for estimating the TN-RN distance\cite{Moore2012,Huang2013,Luo2018,Wang22015,Wang2015,Lin2018}, but they all consider single receiver scenarios. These methods can be summarized as
\begin{itemize}
	\item Use the arrival time of peak concentration (\hspace{-0.04em}\cite{Huang2013,Wang22015}) or the time difference of peak concentration in multiple symbol intervals  to estimate the distance(\hspace{-0.04em}\cite{Moore2012,Luo2018,Lin2018}). However, due to the random walk of molecules, the estimated peak time is usually not accurate.
    \item Use the cumulative number of molecules to estimate distance (\hspace{-0.04em}\cite{Wang2015}). Periodic data is used for iterations which is computationally intensive.
\end{itemize}

The above methods are based on \eqref{eqn:hitting} and \eqref{eqn:CDF}, respectively. However, since the partial differential equation has no closed-form  solution in  SIMO-MCvD systems, the above methods are not applicable here.

In this work, we use the curve fitting method to get the approximate expression for the cumulative number of received molecules. The optimization problem for obtaining $d_k$ and $a_k$ is formulated as
\begin{equation}\label{eqn:d_k}
 \min \limits_{a_k, d_k} \sum\limits_{n = 1}^{{N_{\rm{s}}}} {\left\| {\tilde{F_k}({\bf{p}_\text{T}},n{T_s}) - {F_k}_{\text{fit}}({{\bf{p}}}_\text{T},n{T_s})} \right\|_2^2},
\end{equation}
We use the Levenberg-Marquardt algorithm in \cite{Mor1978} to solve problem \eqref{eqn:d_k}, which takes advantage of the steepest descent and Gauss-Newton method. Let $\bm{\beta}=[a_k,d_k]$ and $\bm{r}(\bm{\beta}) = [r_1(\bm{\beta}), ... r_K(\bm{\beta})]^{\text T}$, where residual $\bm{r}({\bm{\beta}})={\tilde{F_k}({\bf{p}_\text{T}},n{T_s}) -{F_k}_{\text{fit}}({{\bf{p}}}_\text{T},n{T_s})}$.
The LM algorithm iteratively updates the value of the estimation of $\bm{\beta}$ by solving

\begin{equation}\label{eqn:diedai}
\big[{{\bf{J}({\bm{\beta}})}^{\text{T}}}{\bf{J}({\bm{\beta}})} + \mu {\bf{I}}\big]\delta_{\bm{\beta}}= -{{{\bf{J}}({\bm{\beta}})}^{\text{T}}} \bm{r}({\bm{\beta}}),
\end{equation}
where $\delta_{\bm{\beta}}=\bm{\beta}^{(s+1)}-\bm{\beta}^{(s)}$ and ${\bf J }\in \mathbb R^{2}$ is the Jacobian matrix and its element is given by ${\bf{J}}_{ij}=\partial \bm{r}_i  /\partial {\bm{\beta}}_j$, where $i = 1,...,K$ and $j = 1,2$. The details are summarized in Table \ref{tab:LM}.

\begin{table}[t]
	\centering
	\caption{LM-based Distance Estimation (LM-DE)}
	\label{tab:LM}
	\vspace*{-0.2cm}
	\renewcommand\arraystretch{1.5}

	\begin{tabular}{l}
		\hline
		\hline
		$\,\;$1: \textbf{Input:}  $F_k$, $Q$, $D$, $r_k$, $K$,$\varepsilon$, $v$ and the locations of RNs                \\
		$\,\;$2: \textbf{Initialization:} $d_k^{(0)}=a_k^{(0)}=0$,
		$\bm{\beta}^{(0)}=[d_k^{(0)},a_k^{(0)}]$\\
		$\,\;$3: \textbf{for} $s=0,1,2,... ,S$ \\
		$\,\;$4: \quad calculate ${\bf{J}}^{(s)}$ and ${\bf{r}}({\bm{\beta}})^{(s)}$\\
		$\,\;$5: \quad use \eqref{eqn:diedai} to calculate $\delta_{\bm{\beta}}^{(s)}$; \\
		$\,\;$6:         \quad \textbf{if} $\delta_{\bm{\beta}}^{(s)} < \varepsilon$, $\bm{\beta}\leftarrow  [d_{k},a_{k}]$;\\
		$\,\;$7:   \quad \textbf{else if}  ${\bf{r}}({\bm{\beta}}+\delta_{{\bm{\beta}}})^{(s+1)} < {\bf{r}}({{{\bm{\beta}}}})^{(s)}$; \\
		$\,\;$8:    \quad\quad\quad\quad  ${\bm{\beta}}^{(s+1)}\leftarrow{\bm{\beta}}^{(s)}+\delta_{\bm{\beta}}^{(s)}$, $\mu  \leftarrow {\mu  \mathord{\left/{\vphantom {\mu  v}} \right.\kern-\nulldelimiterspace} v}$, return 4;\\	
		$\,\;$9: \quad\quad \textbf{else} $\mu  \leftarrow \mu v$, return 4; \\
		10:         \quad \textbf{end if}\\
		11: \textbf{end for}\\
		12: \textbf{Output:} $\bf{\beta}$\\
		\hline
	\end{tabular}
	\vspace*{-0.2cm}
\end{table}

\begin{table}[t]
	\centering
	\caption{Fitting Parameters and Results}\label{tab:fitting}
	\vspace*{-0.2cm}
	\renewcommand\arraystretch{1.5}
	\begin{tabular}{cccccccc}
		\hline
		\hline
		Parameters  & $a_k$   &$\delta_d$  &  R-square&  SSE  \\
		\hline
		RN1         & 0.6796  &0.0216      &  0.9947  &  5.4823e-04      \\
		\hline
		RN2         &0.5299   &0.0679      &  0.9984  &  3.2167e-04     \\
		\hline
		RN3         &0.6781   &0.0203      &  0.9987  &   6.2183e-04    \\
		\hline
		RN4         &0.5310   &0.0670      &  0.9988  &   2.6105e-04   \\
		\hline
	\end{tabular}
	\vspace*{-0.5cm} 	
	
\end{table}

\begin{figure}[t] 
	\centering  
	\vspace{-0.35cm} 
	\subfigtopskip=2pt 
	\subfigure[]{		
		\label{fitting.sub.1}		
		\includegraphics[width=0.48\linewidth]{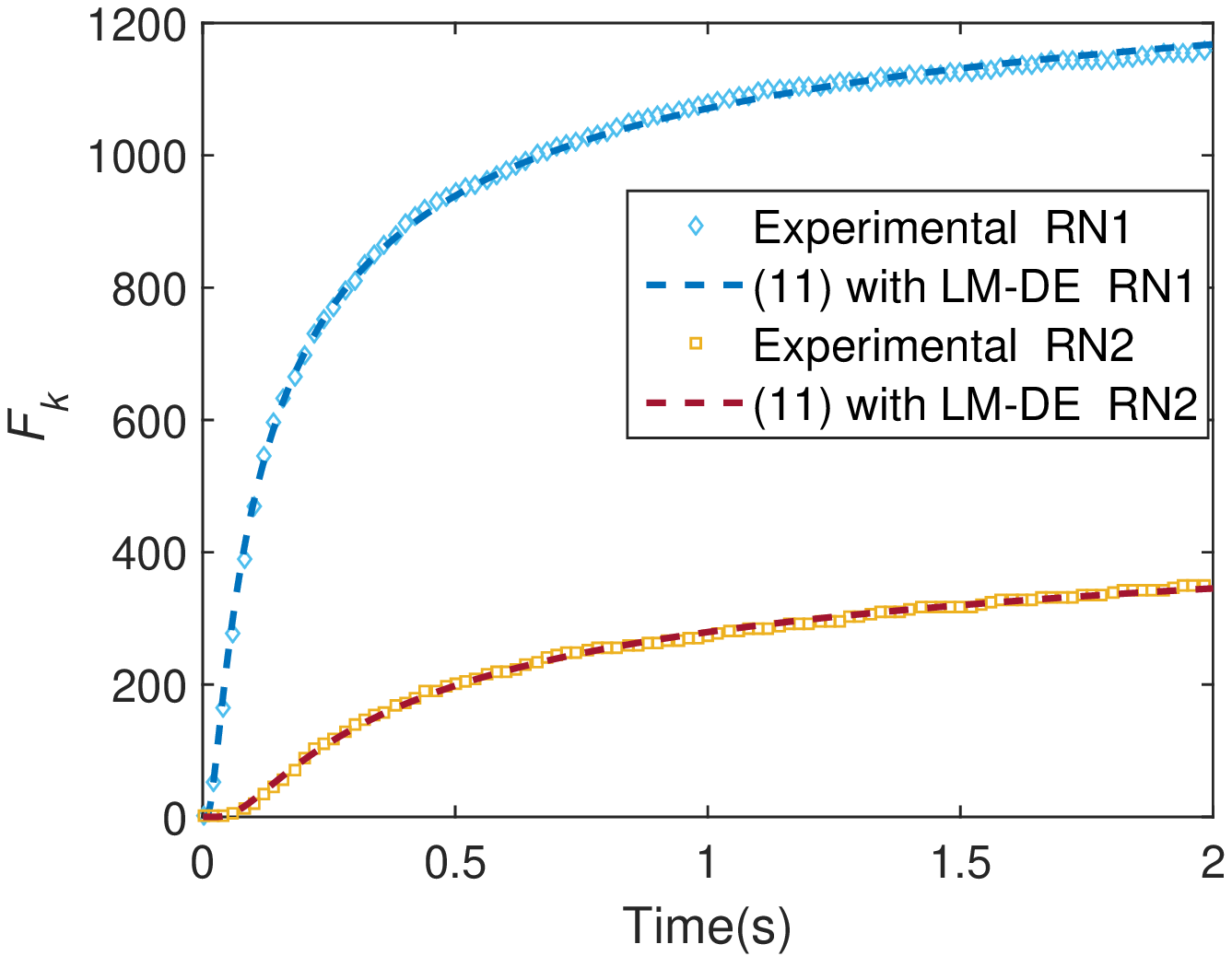}}	
	\subfigure[]{		
		\label{fitting.sub.2}		
		\includegraphics[width=0.48\linewidth]{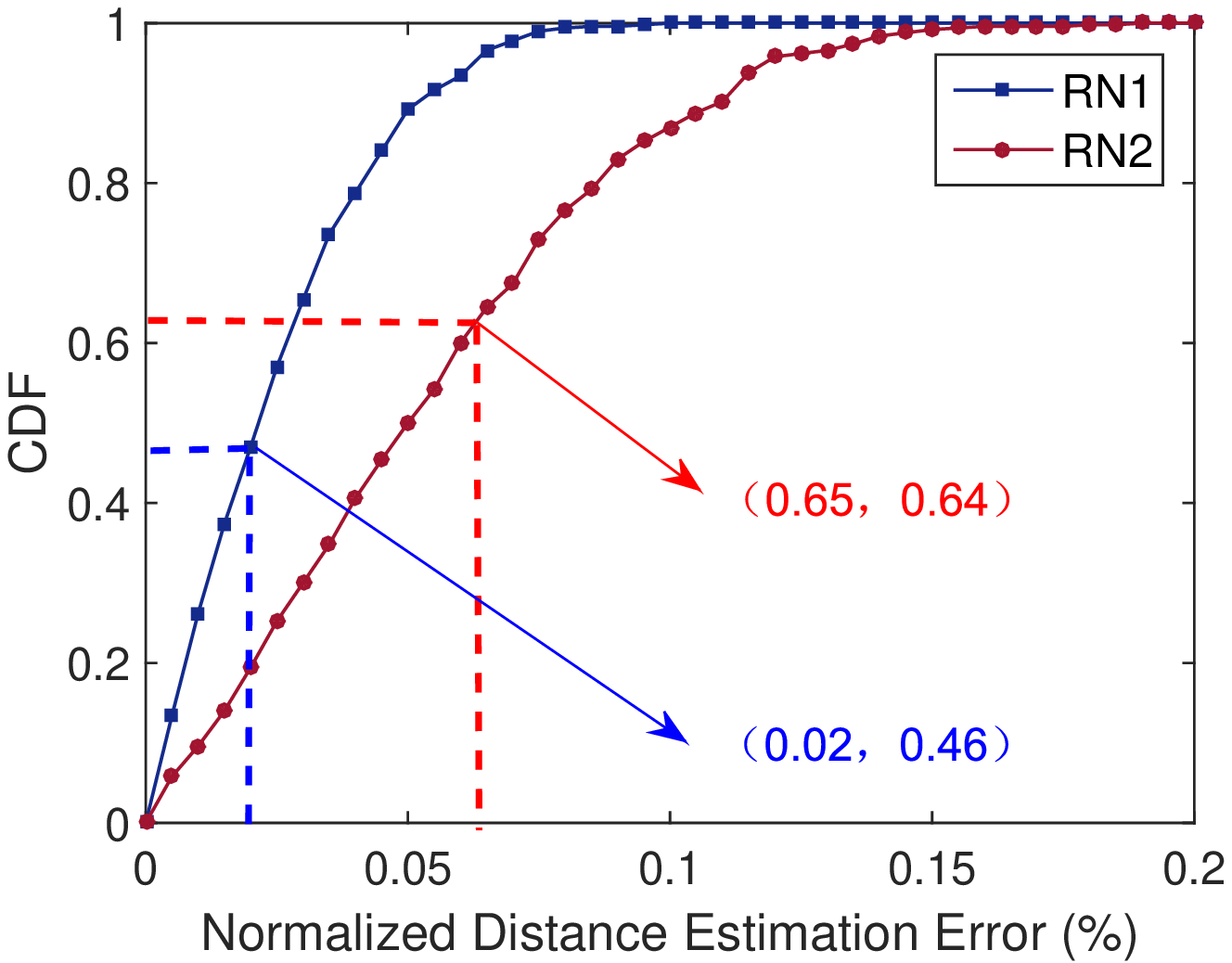}}	  	
    \subfigure[]{		
		\label{fitting.sub.3}		
		\includegraphics[width=.7\linewidth]{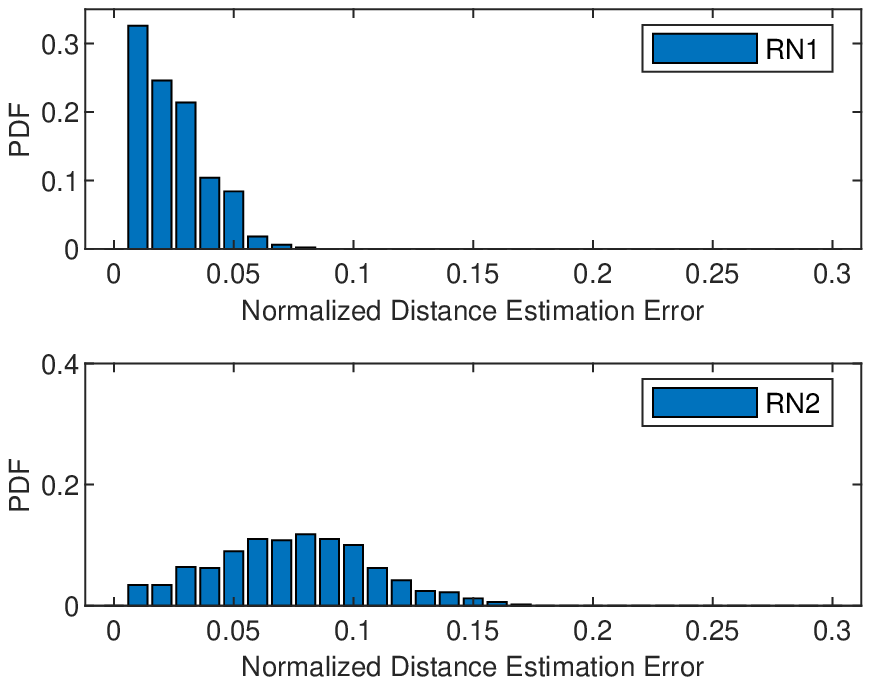}}	  	
	\caption{(a) shows the simulation results of $F_k$ and those using LM-DE. (b) and (c) show the CDF and PDF of the normalized distance estimation error, respectively.  }	
	\label{fig:4RN}
	\vspace{-0.35cm}
\end{figure}

 Table \ref{tab:fitting} and Fig. \ref{fig:4RN} show the accuracy of the proposed LM-DE algorithm, where TN is located at (0,0,0); $\text{RN}_1$, $\text{RN}_2$, $\text{RN}_3$ and $\text{RN}_4$ are located at (0,5,0), (0,0,10), (0,-5,0) and (10,0,0), respectively; $Q=10000$, $D$=100$\upmu$m$^2$/s, $r_k=1\upmu$m. In Table \ref{tab:fitting}, SSE is the sum of squared fitting error and R-square is the coefficient of determination which measures the ability of the fitting function to interpret the data. SSE are within the acceptable range and R-square is close to 1, indicating the fitting is accurate.  The dotted lines in \reffig{fitting.sub.1} represent for the simulation results and the dashed lines are those using LM-DE, where we run the particle-based simulation once randomly.  Only results for $\text{RN}_1$ and $\text{RN}_2$ are given due to symmetry.  It can be seen that, the curves obtained using LM-DE coincide with the simulation results, which demonstrate their accuracy. 

The normalized distance estimation error $\delta_{d_k} \triangleq \left |  \frac{\hat d_k - d_k}{d_k} \right |$ in Table \ref{tab:fitting} is used to measure the accuracy of target parameter $d_k$. We run the simulation 500 times and the average errors are around $2\%$ and $6\%$ for $\text{RN}_1$ and $\text{RN}_2$ , respectively. In \reffig{fitting.sub.2} and \reffig{fitting.sub.3}, the Cumulative Distribution Function(CDF) and the Probability Density Function (PDF) of $\delta_k$ are shown. LM algorithm can estimate the distance from TN to each RN with a reasonable error. Most of the estimation errors are within 10\% for $\text{RN}_2$ and 5\% for $\text{RN}_1$ since the latter is closer to the TN.

\begin{figure}[t]
	\centering
	\includegraphics[width=0.95\linewidth]{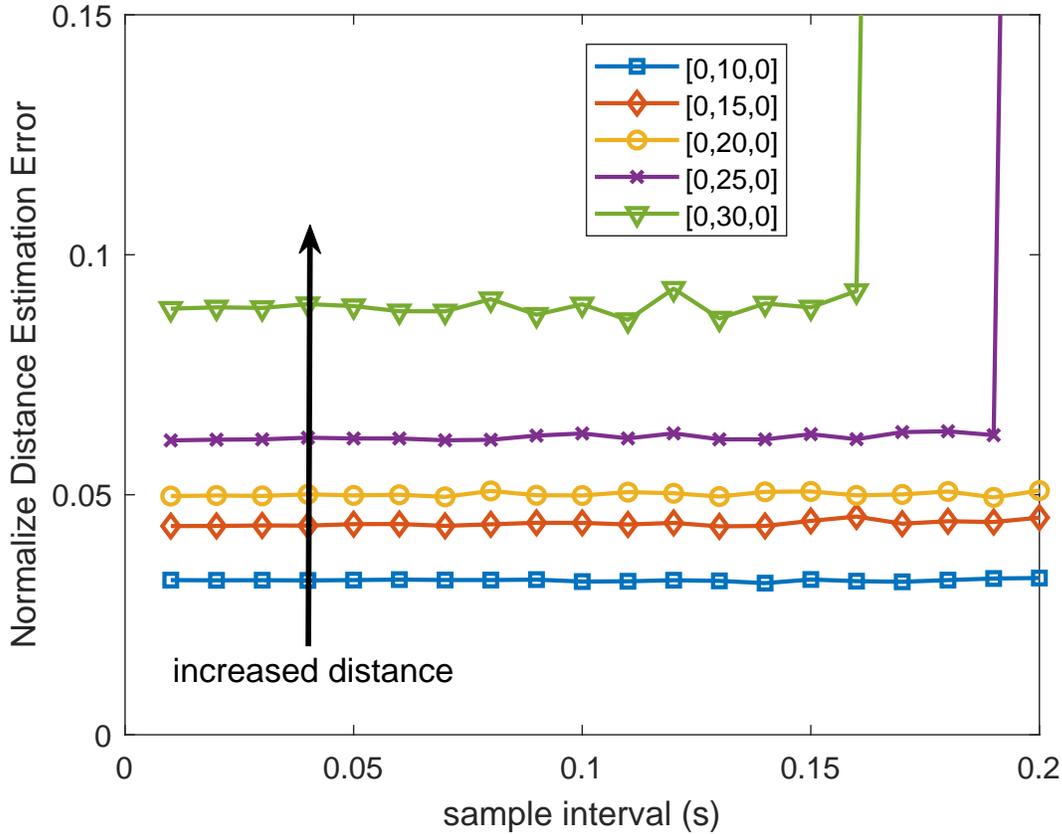}
	\caption{The normalized distance estimation error of RN1 versus sample interval  for different location,  $r_k=1\upmu$m, $Q=10000$ molecules, $D=100\upmu$m$^2$/s. }\label{fig:interval_r1}
\end{figure}

The sampled data is used in the LM method to fit the curve. The number of samples determines the accuracy of distances and the complexity of the algorithm. In order to reduce computational complexity, we need to find the minimum number of samples that are required for estimation, i.e., the maximum sample interval. Fig. \ref{fig:interval_r1} shows normalized estimation error of different sample interval and TN's location where the TN is located at (0,5,0), (0,10,0), (0,15,0), (0,20,0), (0,25,0) respectively for five experiments; other parameters are the same as Fig. \ref{fig:4RN}. It can be seen that when the sample interval is less than a certain threshold, the estimation error is maintained at a certain level. The threshold is the maximum sample interval, and then further increase the number of samples, does not improve the estimation performance. Finding the threshold can reduce the number of samples without sacrificing accuracy, which help to reduce the complexity.

\subsection{Source Localization}
In the previous section, the distance from TN to $\text{RN}_k$ is obtained. The problem of finding the estimated coordinates of TN is formulated as
\begin{equation}\label{eqn:p_t}
\begin{split}
{\bf{p}}_{\text{T}}^* &= \arg \min\limits_{\bf{p}_\text{T}} \sum\limits_{k = 1}^K { \big(  {{{\left\| {{\bf{p}}_\text{T} - {{\bf{p}}_k}} \right\|}_2^2} - {d_k}^2 }\big) }^2\\
&\triangleq  \arg \min\limits_{\bf{p}_\text{T}} H.
\end{split}
\end{equation}

Note that \eqref{eqn:p_t} is not convex in general and it is very difficult to obtain closed-form solution. Therefore, we propose to use steepest descent method to find a locally optimal solution to \eqref{eqn:p_t}. The steepest descent method takes the derivative of the objective function as the negative gradient direction, and uses a linear search method to calculate the iteration step size which has fast convergence and less calculation. The performance of the steepest descent algorithm relies on the initial point which can be obtained by multi-point localization method. Assuming $d_k$'s are accurate, it is easy to obtain that  the coordinates of TN satisfy
 \begin{equation}\label{eqn:position}
\left\{
\begin{aligned}
{{({x_{\text{T}}} - {x_1})}^2} + {{({y_{\text{T}}} - {y_1})}^2} + {{({z_{\text{T}}} - {z_1})}^2} &  = {d_1}^2\\
{{({x_{\text{T}}} - {x_2})}^2} + {{({y_{\text{T}}} - {y_2})}^2} + {{({z_{\text{T}}} - {z_2})}^2} &  = {d_2}^2\\
 \cdots \\
{{({x_{\text{T}}} - {x_K})}^2} + {{({y_{\text{T}}} - {y_K})}^2} + {{({z_{\text{T}}} - {z_K})}^2} &  = {d_K}^2.
\end{aligned}
\right.
\end{equation}

It is worth noting that \eqref{eqn:position} has a unique solution only when $K \ge 4$, i.e., the number of RN must be no less than four for the 3D  scenarios considered here. Solving \eqref{eqn:position} based on least square criterion yields

\begin{equation}\label{eqn:posi}
{{\bf{p}}_{\text{T}}^{(0)}}= \big[({\bf{A}}^{\text{T}} {\bf{A}})^{-1} {{\bf{A}}^{\text{T}}}{\bf{b}}\big]^{\text{T}},
\end{equation}
in which,
\begin{equation}
{\bf{A}} = \left[ {\begin{array}{*{20}{c}}
{2({x_K} - {x_1})}\\
{2({x_K} - {x_2})}\\
 \vdots \\
{2({x_K} - {x_{K - 1}})}
\end{array}\begin{array}{*{20}{c}}
{2({y_K} - {y_1})}\\
{2({y_K} - {y_2})}\\
 \vdots \\
{2({y_K} - {y_{K - 1}})}
\end{array}\begin{array}{*{20}{c}}
{2({z_K} - {z_1})}\\
{2({z_K} - {z_2})}\\
 \vdots \\
{2({z_K} - {z_{K - 1}})}
\end{array}} \right],
\end{equation}
and
 \begin{equation}
{\bf{b}} = \left[ {\begin{array}{*{20}{c}}
{ - M_1 + M_K + {d_1}^2 - {d_K}^2}\\
{ - M_2 + M_K + {d_2}^2 - {d_K}^2}\\
 \vdots \\
{ - M_{K-1} + M_K + {d_{K - 1}}^2 - {d_K}^2}
\end{array}} \right],
\end{equation}
where $M_k={x_k}^2 + {y_k}^2 + {z_k}^2$.

\begin{table}[t]
	\centering
	\caption{The Proposed CSL Method }
	\label{tab:CSL}
	\vspace*{-0.2cm}
	\renewcommand\arraystretch{1.5}
	\begin{tabular}{l}
		\hline
		\hline
		$\,\;$1: \textbf{Input:}  $F_k$, $Q$, $D$, $r_k$, $K$, $\lambda=1/2$ and the locations of RNs                \\
		$\,\;$2: \textbf{for} $k = 1, \cdots ,K$  ($K \ge 4$) \\
		$\,\;$3: \quad use LM-DE method to estimate parameter $a_k$ and $d_k$\\
		$\,\;$4: \textbf{end for}\\		
		$\,\;$5: \textbf{Initialization:} compute ${\bf{p}}_{\text{T}}^{(0)} $ according to \eqref{eqn:posi}\\
        $\,\;$6:  Update ${\bf{p}}_{\text{T}}^{(0)}$ using steepest descent method, until convergence.\\
		$\,\;$6: \textbf{Output:}  ${\bf{p}}_{\text{T}}^*=({x_{\text{T}}^*},{y_{\text{T}}^*},{z_{\text{T}}^*})$\\
		\hline
	\end{tabular}
	\vspace*{-0.2cm}
\end{table}

Take ${{\bf{p}}_{\text{T}}^{(0)}}$ as the initial value and apply the steepest descent (SD) algorithm to obtain TN's location. The proposed CSL method is summarized in Table \ref{tab:CSL}.  Note that it is not always helpful to increase the number of RNs due to the inter-RN interferences, as will be shown in the numerical results.

\subsection{Computational Complexity}

 The complexity of proposed CSL method mainly comes from LM-DE algorithm and the SD algorithm, which have complexity of   ${\mathcal{O}}(KS)$ and ${\mathcal{O}}(S^{'}M)$, respectively. Therefore, the complexity of the CSL method is $\mathcal{O}(KS+S^{'}M)$. ($S,S^{'},M$  are the number of iterations in the algorithm.)

\begin{figure}[t] 
	\centering  
	\subfigtopskip=2pt 
	\subfigcapskip=-5pt 
	\subfigure[]{		
		\label{iteration.sub.1}		
		\includegraphics[width=0.48\linewidth]{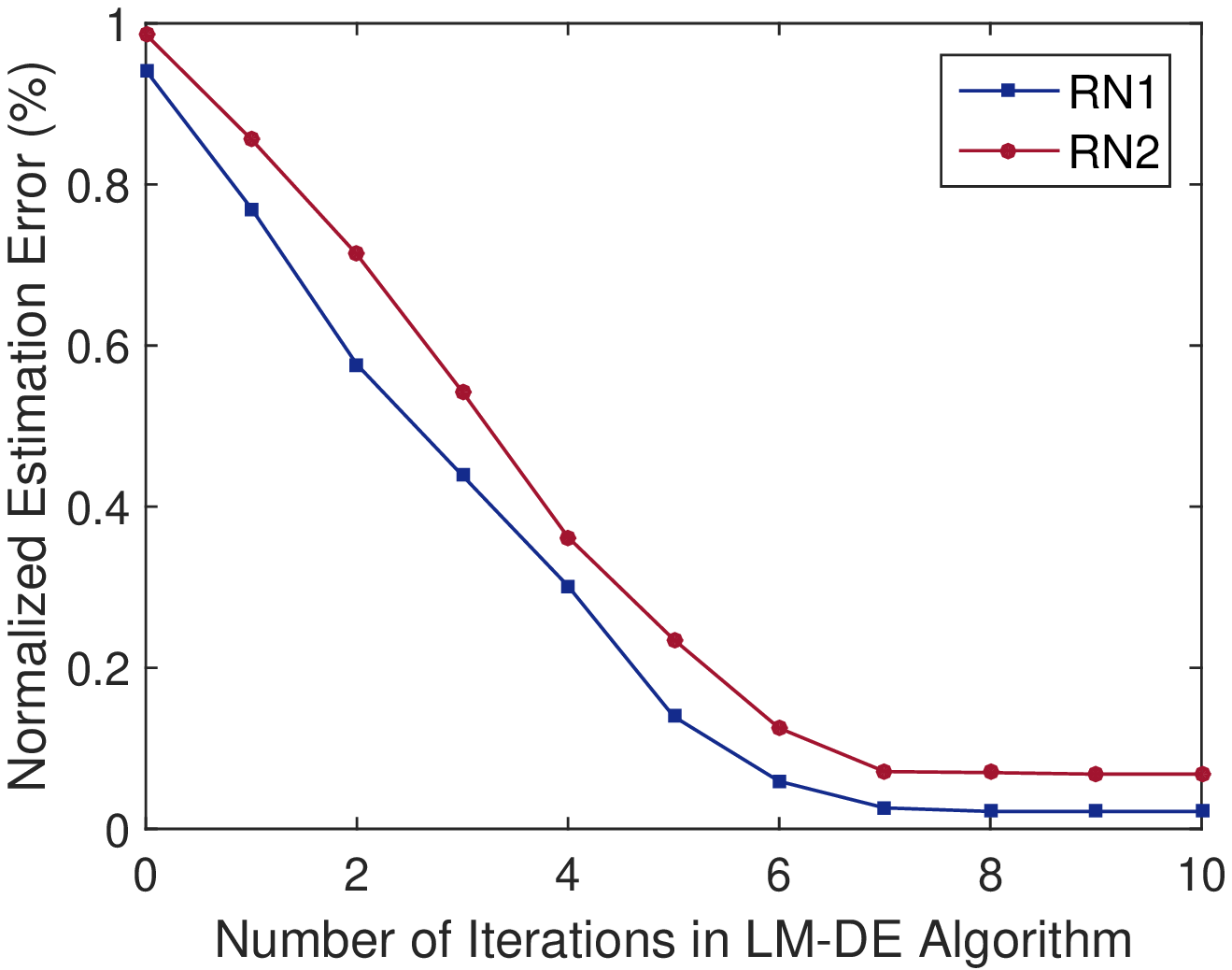}}	
	\subfigure[]{		
		\label{iteration.sub.2}		
		\includegraphics[width=0.48\linewidth]{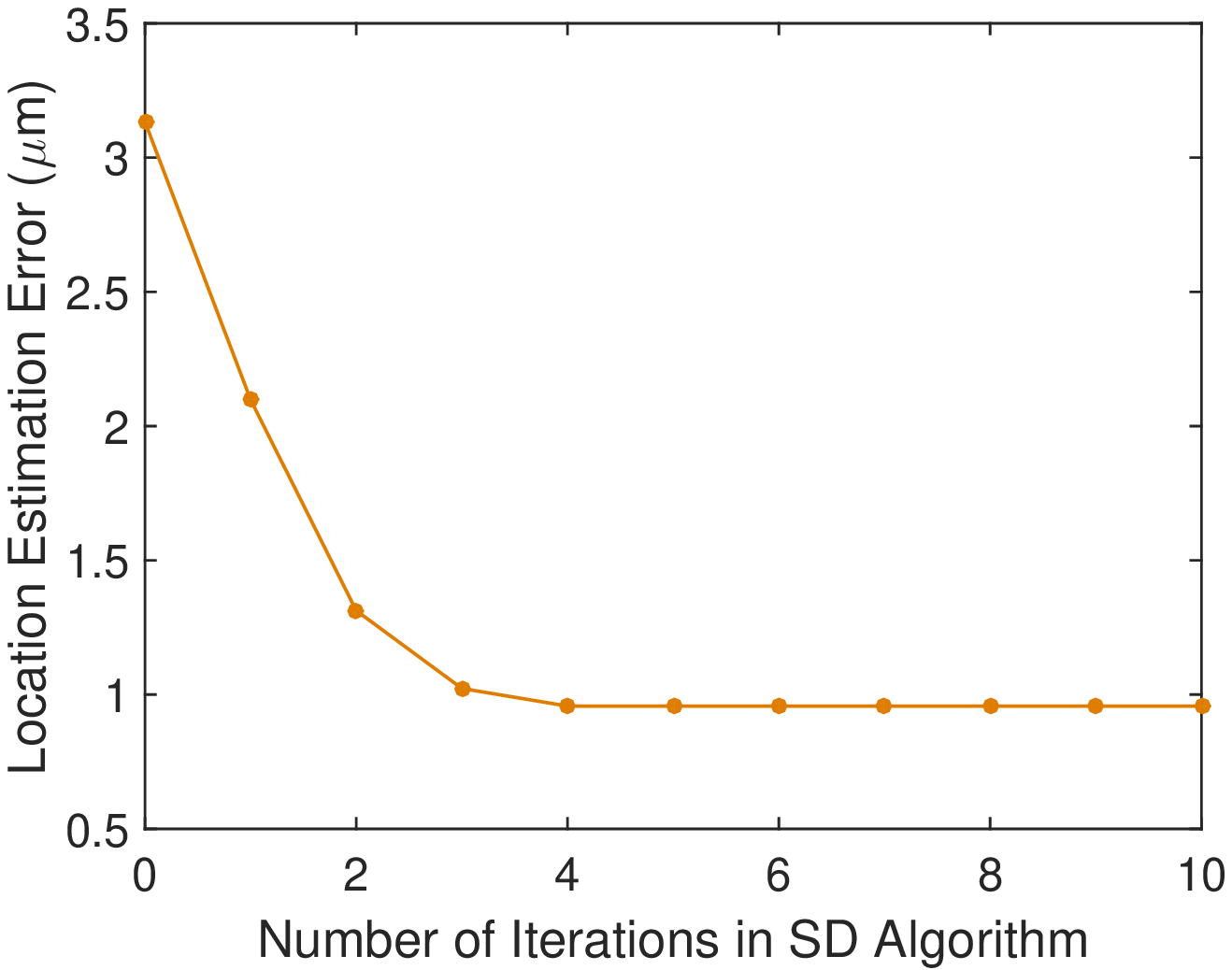}}	  	
	\caption{(a) Normalized distance estimation error versus iterations in LM-DE algorithm. (b) Location estimation error versus iterations in SD algorithm.}	
	\label{fig:iteration}
	
\end{figure}

Fig. \ref{fig:iteration} shows the error performance of the proposed LM-DE and SD algorithms with respect to the number of iterations.  It can be seen that the proposed algorithms converges within a few iterations.

\section{Numerical Results and Discussions}

 In this section, particle-based simulations are carried out to evaluate the performance of the proposed CSL method. The displacement of a diffusing molecule follows $N(0,2DT_s)$ in each of the three dimensions. When a molecule hits the surface of the absorbing receivers, it is removed from the environment.

\begin{figure}[t]
	\centering
	\includegraphics[width=0.7\linewidth]{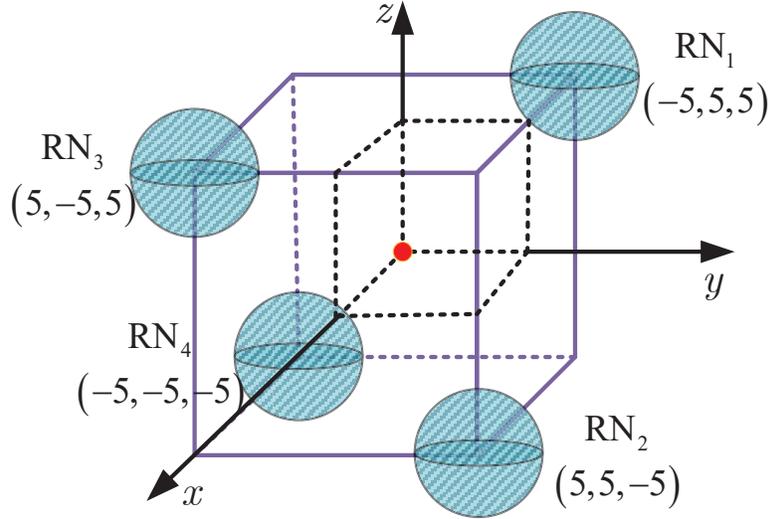}
	\caption{Simulation scenario with one TN and four RNs, where the coordinates of RN1-RN4 are: (-5,5,5), (5,5,-5), (5,-5,5), (-5,-5,-5), respectively.}\label{fig:simulation}
\end{figure}

Unless otherwise stated, the scenario with four RNs located at the four distant vertices of a cube with a side length of 10$\upmu$m is considered in simulation, as shown in Fig. \ref{fig:simulation}. The center of the cube is set to be the coordinate origin and the distances between each RN to the origin are the same.

\begin{figure}[t] 
	\centering  
	\subfigure[RN1]{		
		\label{Cumulative.sub.1}		
		\includegraphics[width=0.45\linewidth]{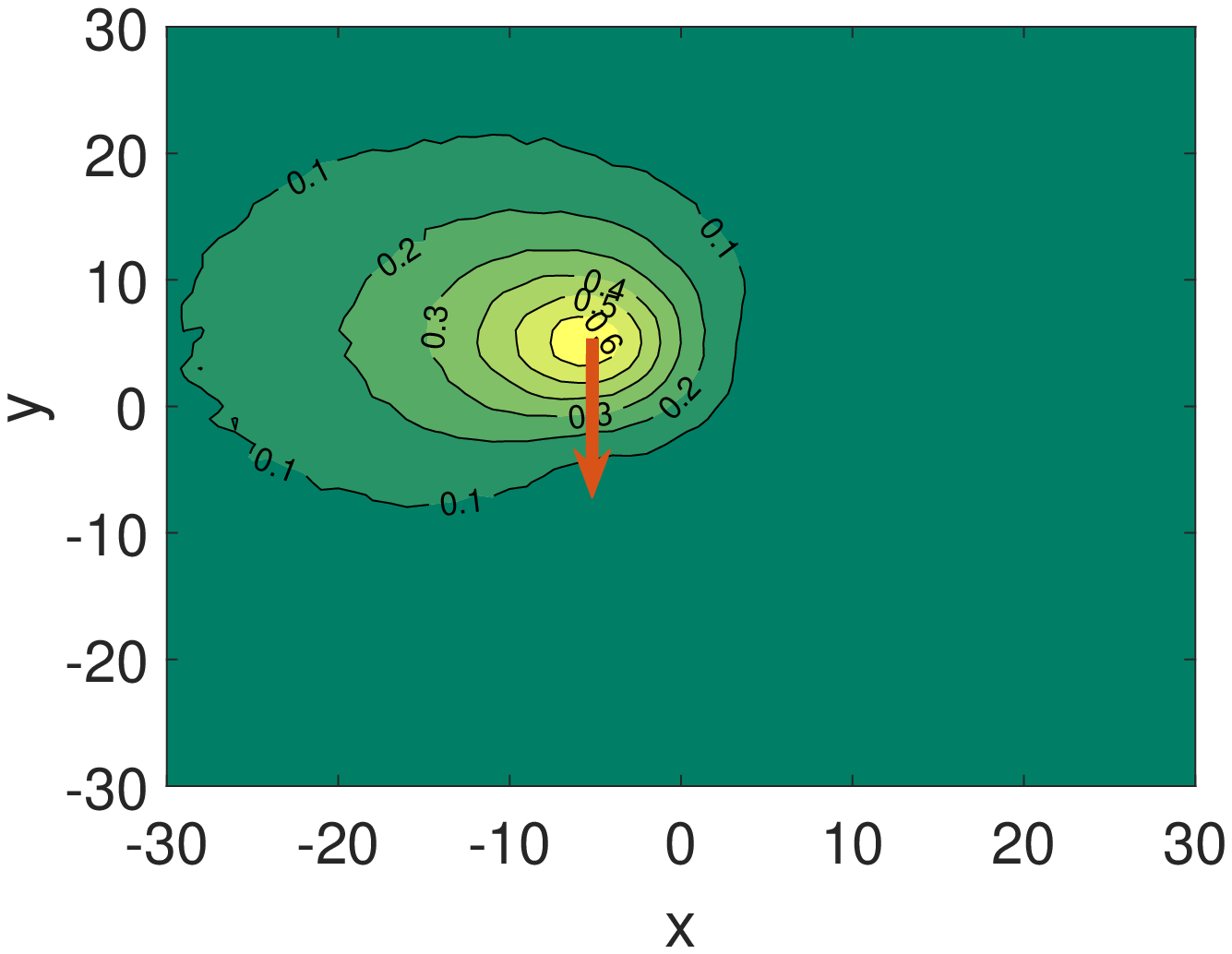}}	
	\subfigure[RN2]{		
		\label{Cumulative.sub.2}		
		\includegraphics[width=0.45\linewidth]{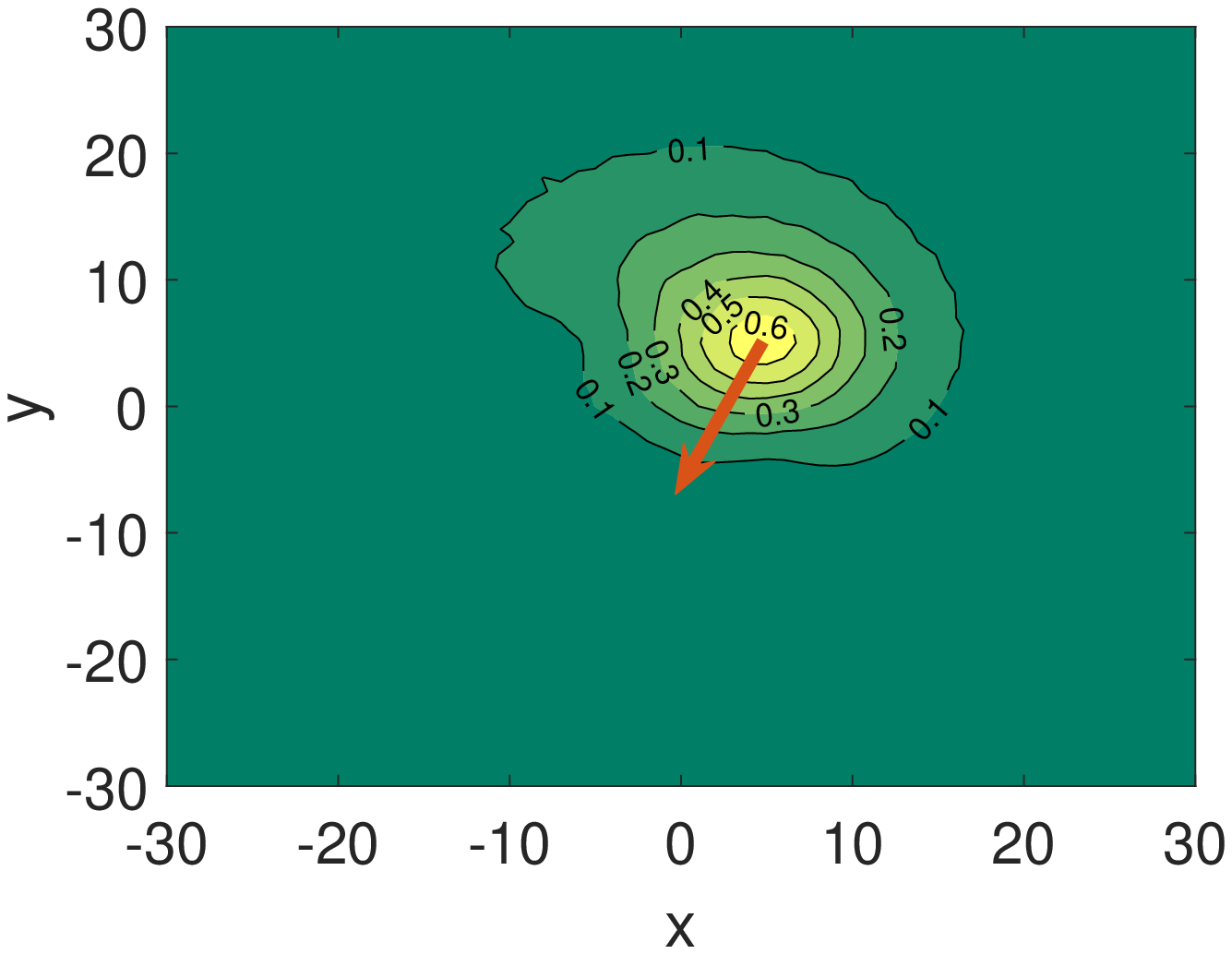}}	  	
	\subfigure[RN4]{		
		\label{Cumulative.sub.3}		
		\includegraphics[width=0.45\linewidth]{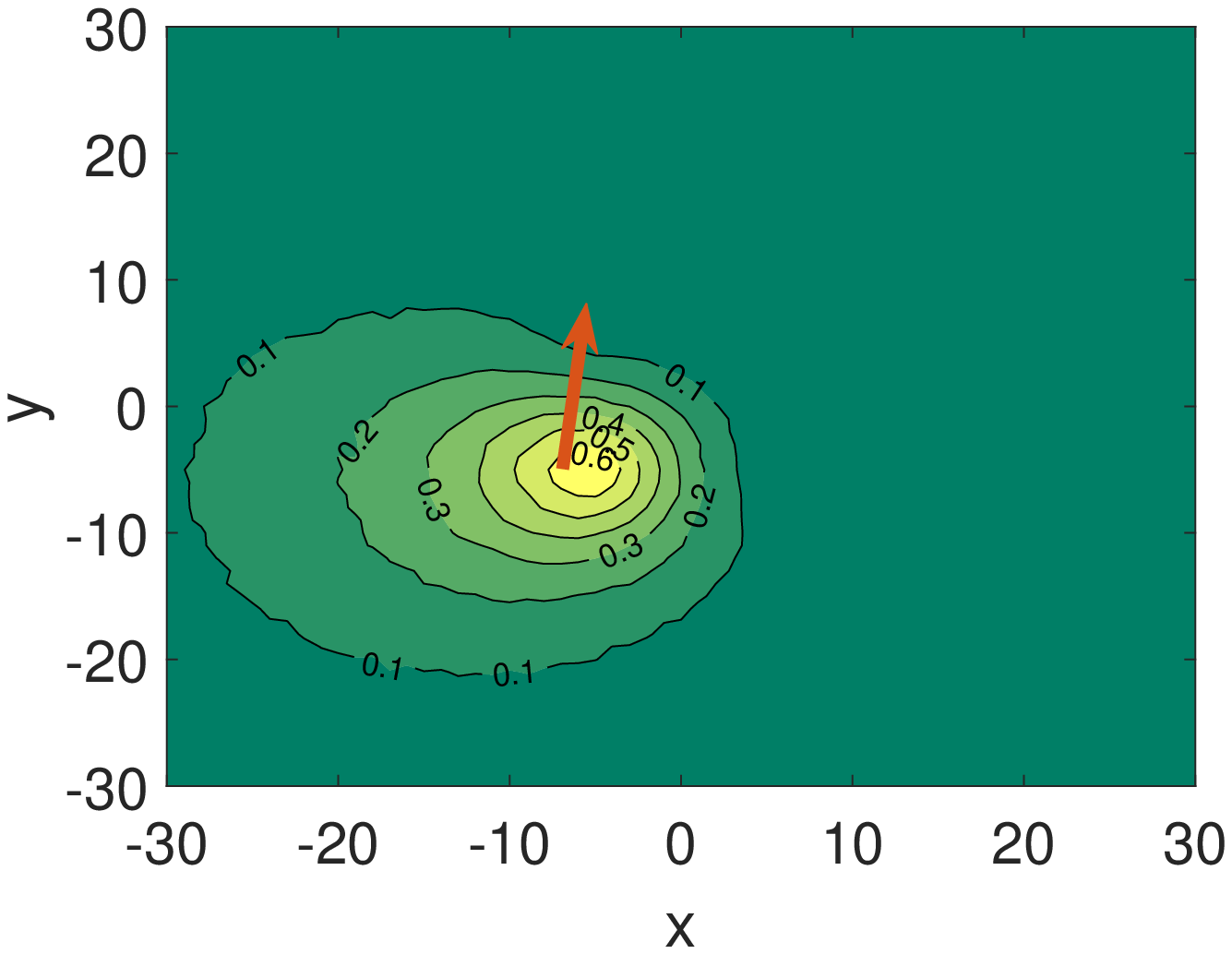}}
	\subfigure[RN3]{		
		\label{Cumulative.sub.4}		
		\includegraphics[width=0.45\linewidth]{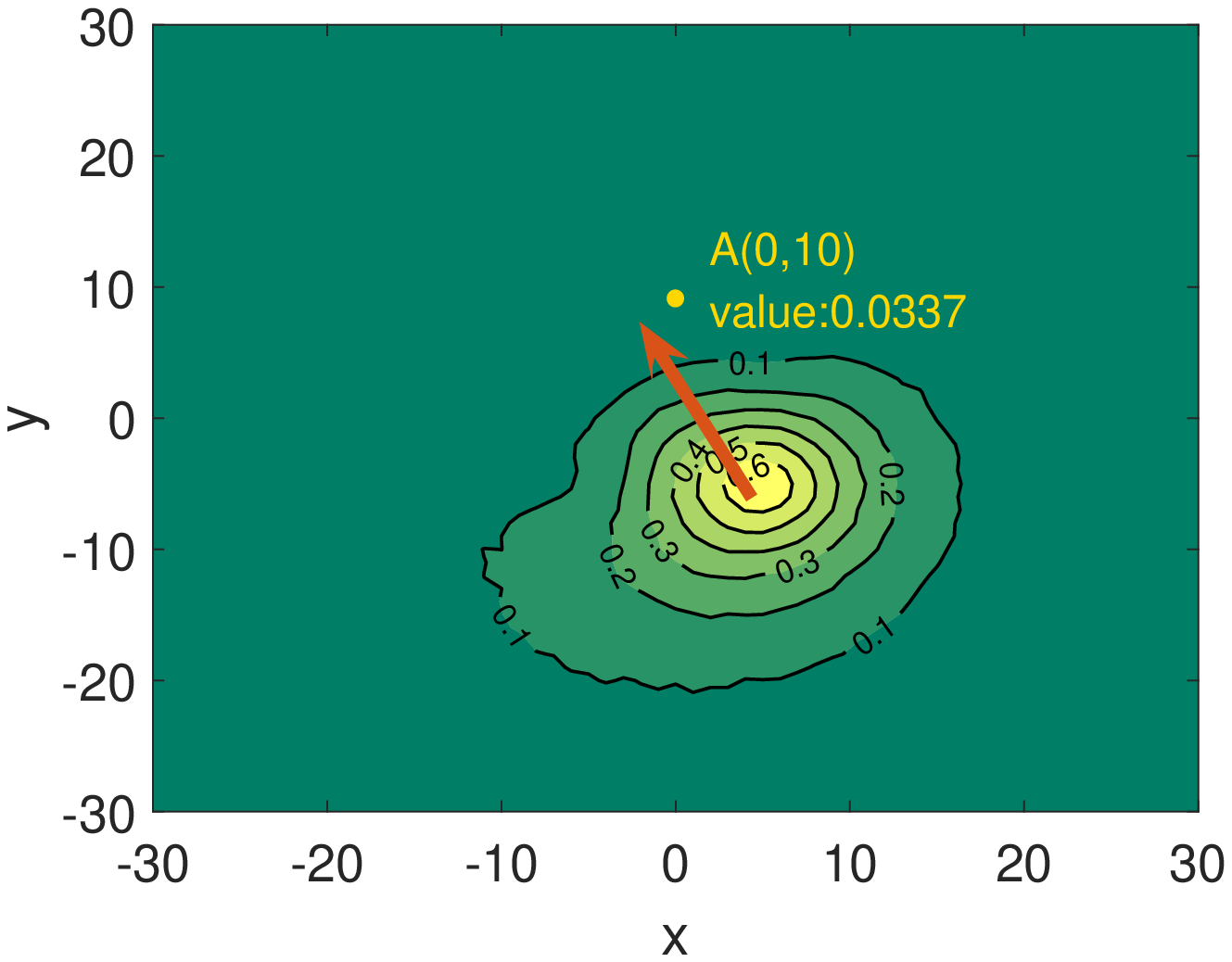}}  	
	\caption{Cumulative probability of received molecules in 2 seconds with different TN's location, where $Q=10000$ molecules, $D=100$ $\upmu$m$^2$/s, $r_k=4$ $\upmu$m.}	
	\label{fig:cul_RN}
\end{figure}

 Fig. \ref{fig:cul_RN}  shows the cumulative probability of receiving molecules within 2 seconds  with respect to different TN's locations in the XoY plane.  For example, Fig. \ref{Cumulative.sub.4} shows that when the TN is located at (0, 10),  the receiving probability for RN3 within 2 seconds is 0.0337. In Fig. \ref{fig:cul_RN}, the red arrow indicates the direction that the receiving probability decrease the most rapidly, which is actually towards other RNs. This means due to the existence of other RNs, the receiving probability shows directivity and this property is related to  the location of RNs and the TN. In general, due to the interplay of RNs, a simple explicit expression for the receiving probability (as well as the channel impulse response) is very difficult to derive.

 \begin{figure}[t]
 	\centering
 	\includegraphics[width=1\linewidth]{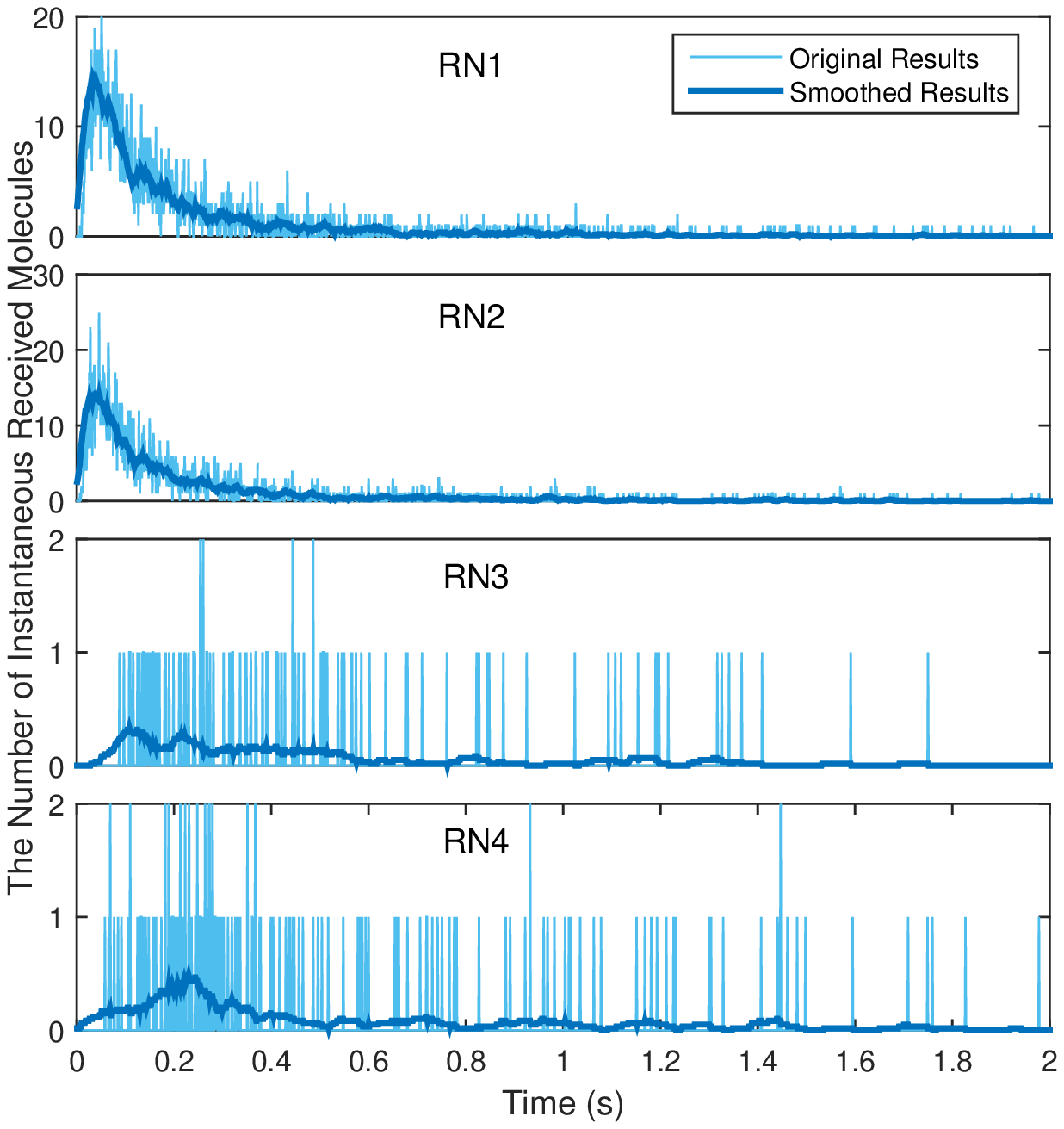}
 	\caption{The instantaneous received molecular number in A, $r_k=4$  $\upmu$m, $Q=10000$ molecules, $D=100$  $\upmu$m$^2$/s, the coordinates of A is ${\bf{p}}_\text{A}=(0,10,0)$.}\label{fig:shunshi}
 \end{figure}

 Fig. \ref{fig:shunshi} shows the instantaneous number of received molecular when TN is located at A(0,10,0). The light-blue line represents the instantaneous number of molecules received by each RN and the dark-blue solid line represents smoothed results using window function. It can be seen that due to the uncertainty of Brownian motion and small quantities of received molecules, it is difficult to find the peak time of received molecules, even for smoothed curves. Since the TN is close to RN1 and RN2, the number of received molecules for RN3 and RN4 varies significantly, which leads to an increase in distance estimation error. Therefore, the schemes in \hspace{-0.04em}\cite{Huang2013,Wang22015,Moore2012,Luo2018} does not work in this case.

\subsection{Performance Evaluation of the Proposed CSL Method}
\begin{figure}[t]
	\centering
    \subfigure[]{
	\includegraphics[width=.45\linewidth]{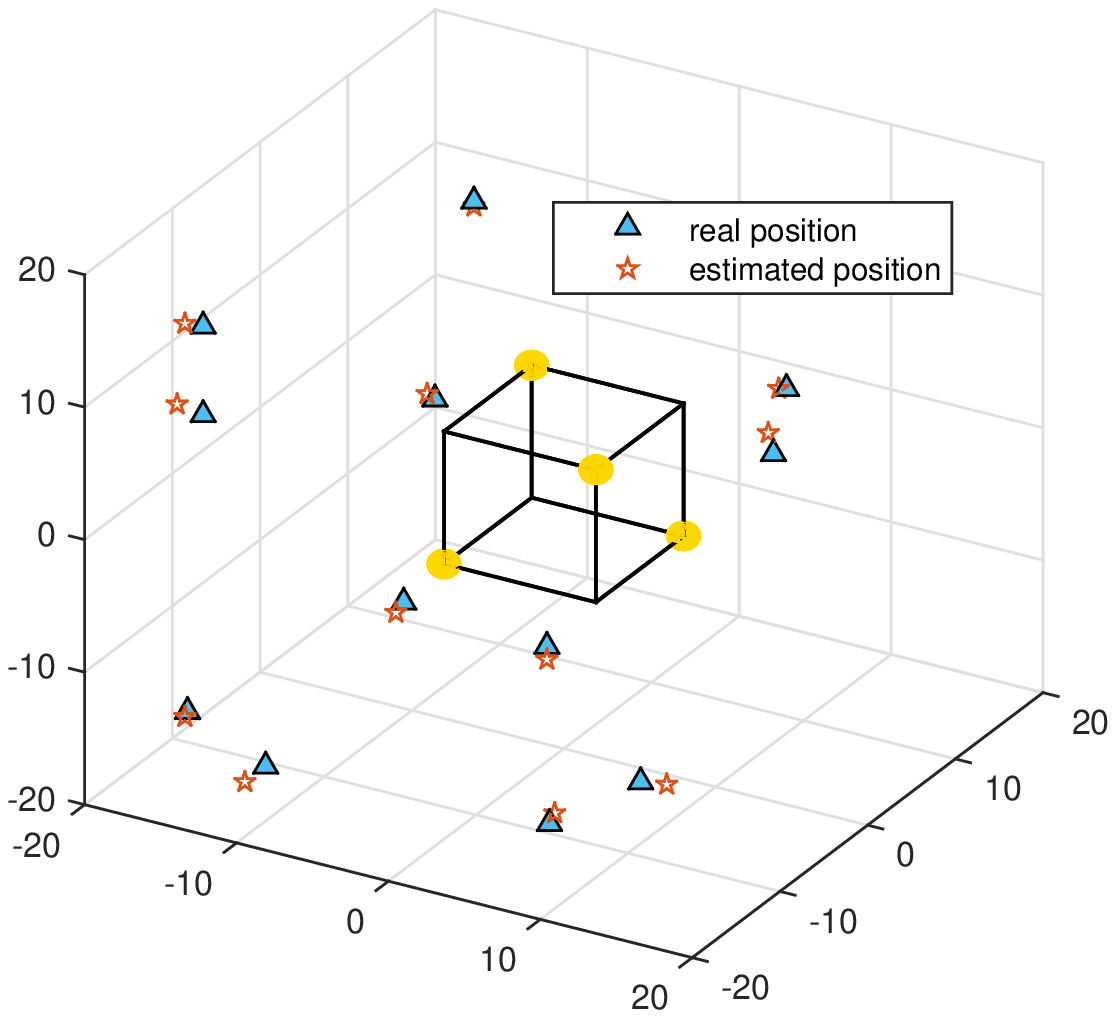}
	\label{fig:position} }
    \subfigure[]{
    \includegraphics[width=.45\linewidth]{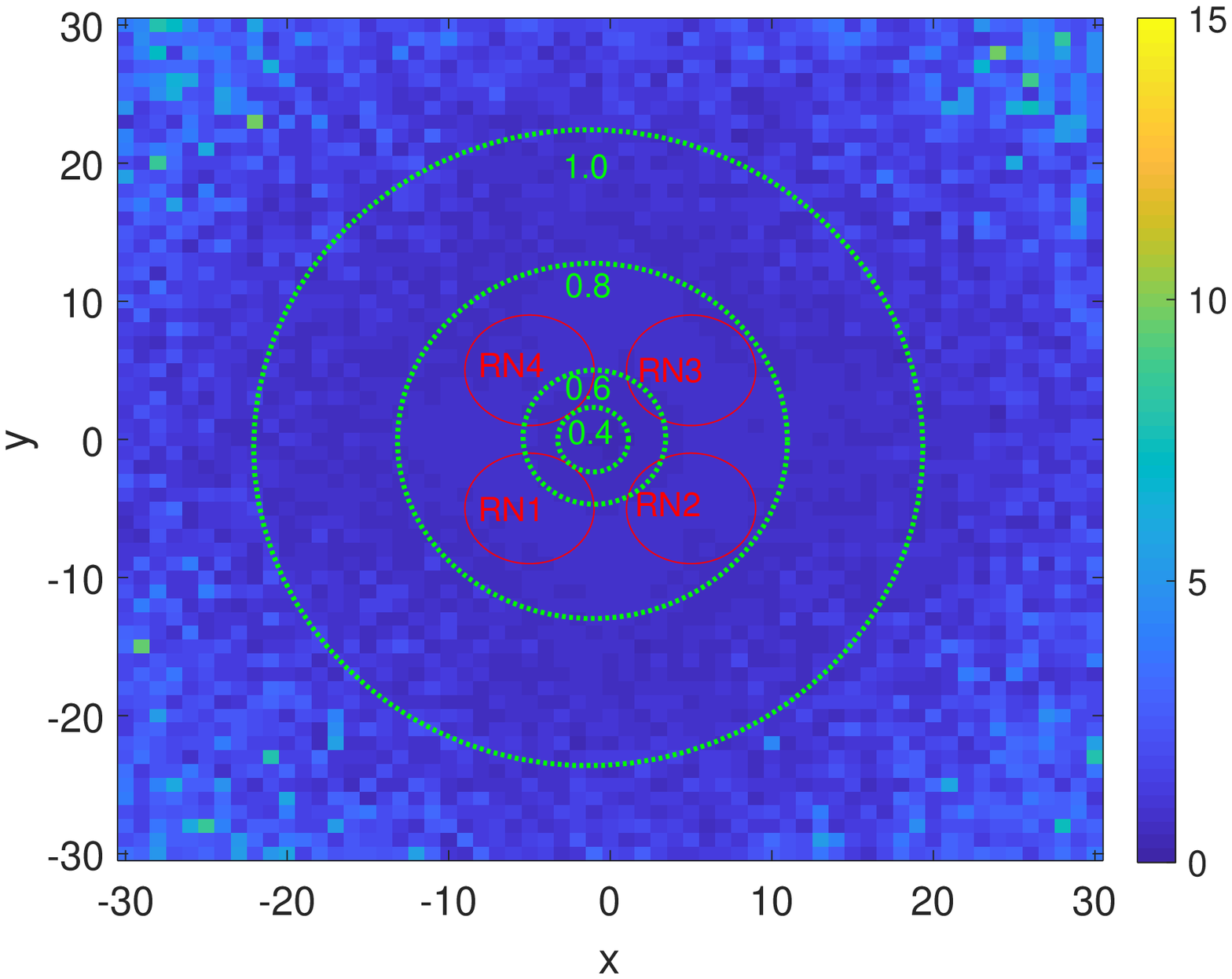}
	\label{fig:100_100_5} }
    \caption{(a) Location estimation error, $r_k=1$ $\upmu$m, $Q=10000$ molecules, $D=100$ $\upmu$m$^2$/s. (b) Location estimation error in the XoY plane, $r_k=4$ $\upmu$m, $Q=10000$ molecules, $D=100$ $\upmu$m$^2$/s. }
\end{figure}

An illustration of the estimation results of CSL for twelve different TN's locations is given in Fig. \ref{fig:position} and Fig. \ref{fig:100_100_5}  where $r_k=1$ $\upmu$m, $Q=10000$ molecules, $D=100$ $\upmu$m$^2$/s and the location estimation  error is defined as
$\delta p = \left\| {{{\bf{p}}_{\text{T}}} - {\bf{p}}_{\text{T}}^*} \right\|_2$. It is clear that the proposed method can accurately estimate the RN's location. In Fig. \ref{fig:100_100_5}, the red circles are the projections of RNs on XoY plane with a radius of 4 $\upmu$m and we show the estimation error for different TN's location in the XoY plane. The green circle marks the approximate range of TN's location with errors less than 0.4 $\upmu$m, 0.6 $\upmu$m, 0.8 $\upmu$m, and 1 $\upmu$m, respectively. It can be seen that in most of the range  $\delta p$ is below 1 $\upmu$m. In general, the farther TN is located from the RNs, the greater the error becomes. Moreover, the error fluctuation is small when the distance is relatively small, while it varies greatly when the distance becomes large. This is because large distance leads to significantly reduced number of received molecules and inaccurate estimation. It is also observed that there are larger errors in the four corner areas, which may be caused by the path blocking effect of multiple RNs. Therefore, the topology of the RNs has a certain impact on the localization performance.

\subsection{Factors that Affects Localization Accuracy}

\begin{figure*}[htb] 
	\centering  
	\vspace{-0.35cm} 
	\subfigtopskip=2pt 
	\subfigcapskip=-5pt 
	\subfigure[The location estimation error versus TN's location  for different receiver radius, $Q=10000$ molecules, $D=100$ $\upmu$m$^2$/s.]{		
		\label{level.sub.1}		
		\includegraphics[width=0.46\linewidth]{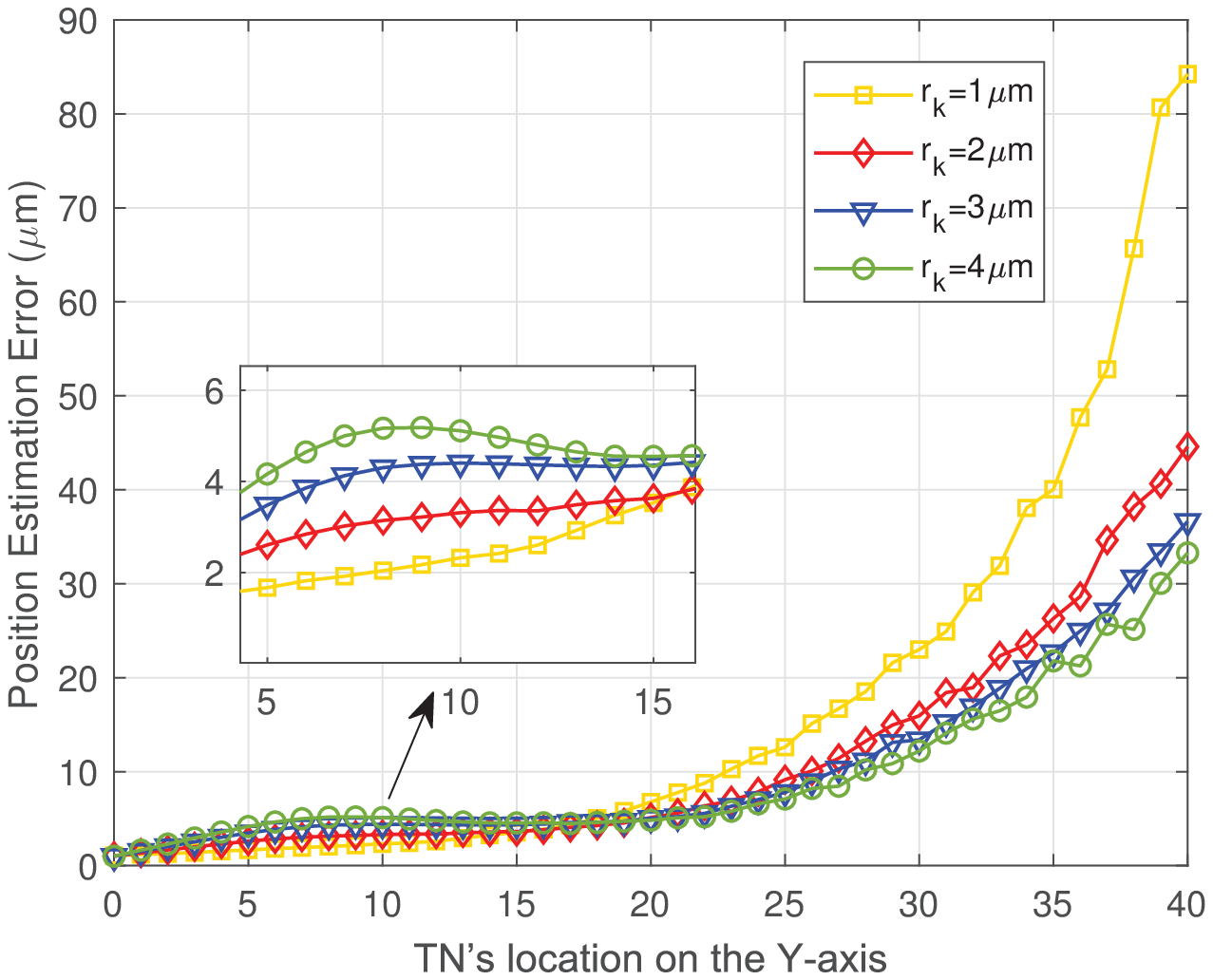}
		
	}	
	\subfigure[The location estimation error versus TN's location  for different number of transmitted molecules, $r_k=1$ $\upmu$m, $D=100$ $\upmu$m$^2$/s.]{		
		\label{level.sub.2}		
		\includegraphics[width=0.46\linewidth]{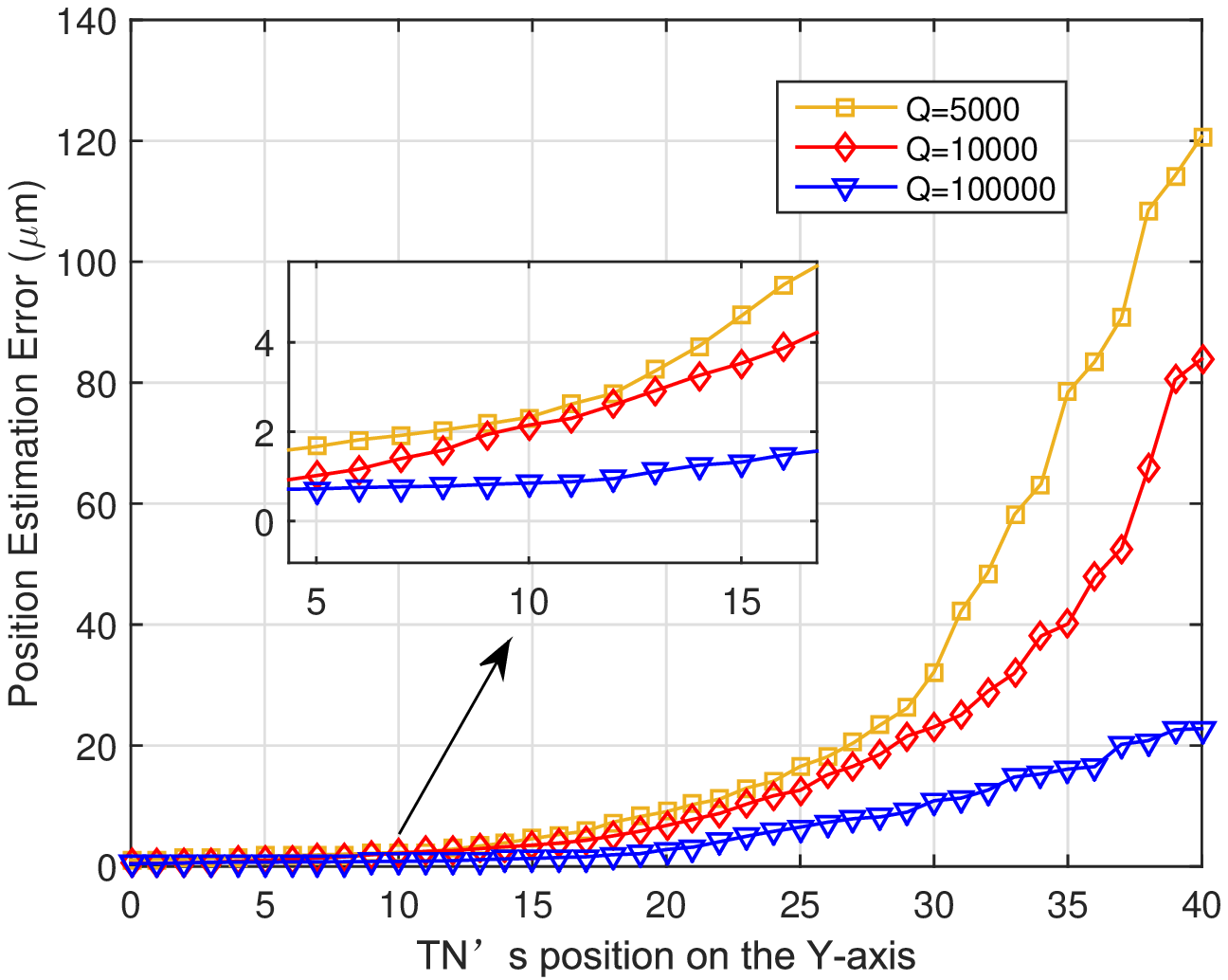}}
	\subfigure[The location estimation error of RN1 versus TN's location  for different diffusion coefficients, $r_k=1$ $\upmu$m, $Q=10000$ molecules.]{		
		\label{level.sub.3}		
		\includegraphics[width=0.46\linewidth]{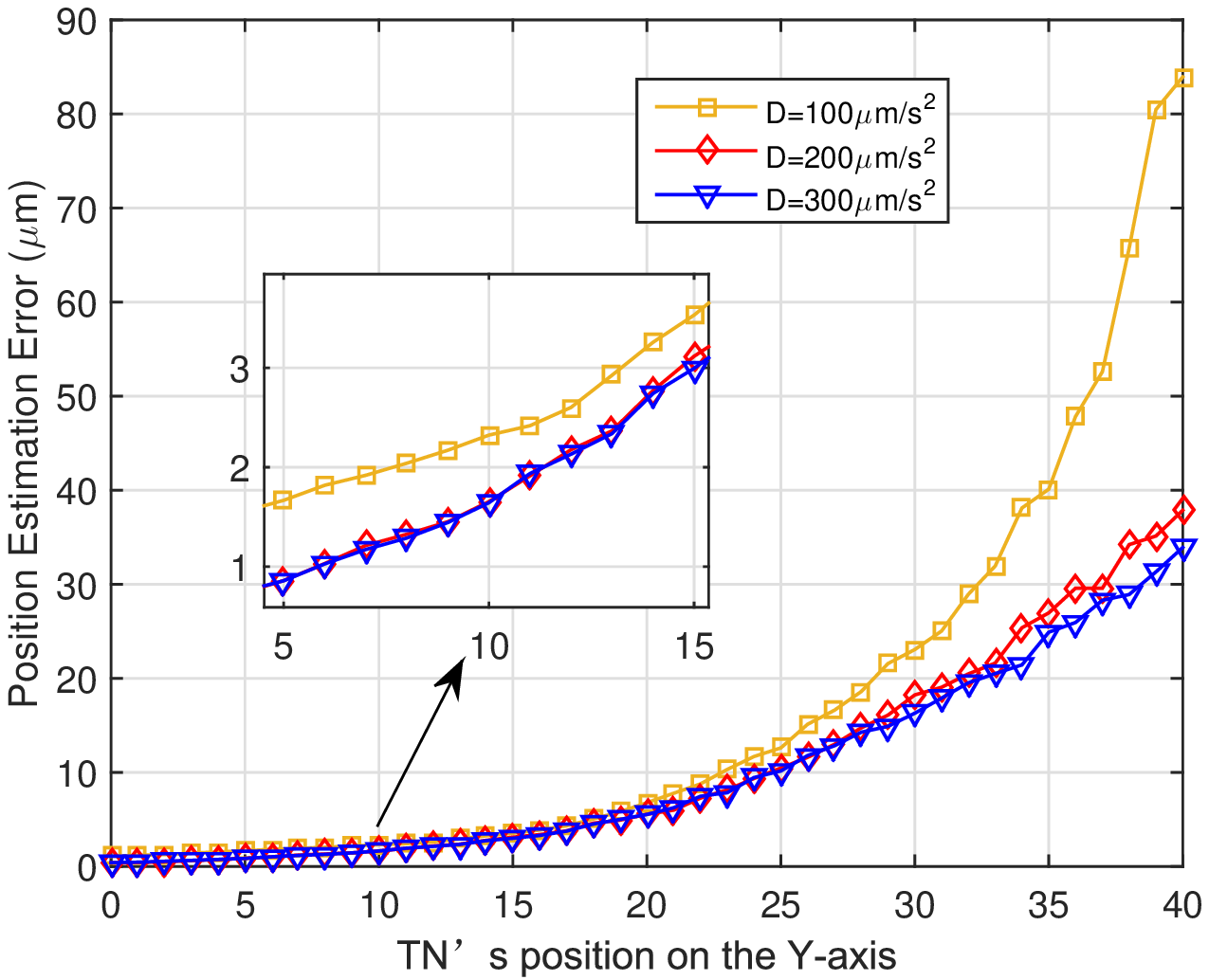}}		
	\subfigure[The location estimation error versus TN's location  for different flow velocity, $r_k=4$ $\upmu$m, $Q=10000$ molecules, $D=100$ $\upmu$m$^2$/s.]{		
		\label{level.sub.4}
		\includegraphics[width=0.46\linewidth]{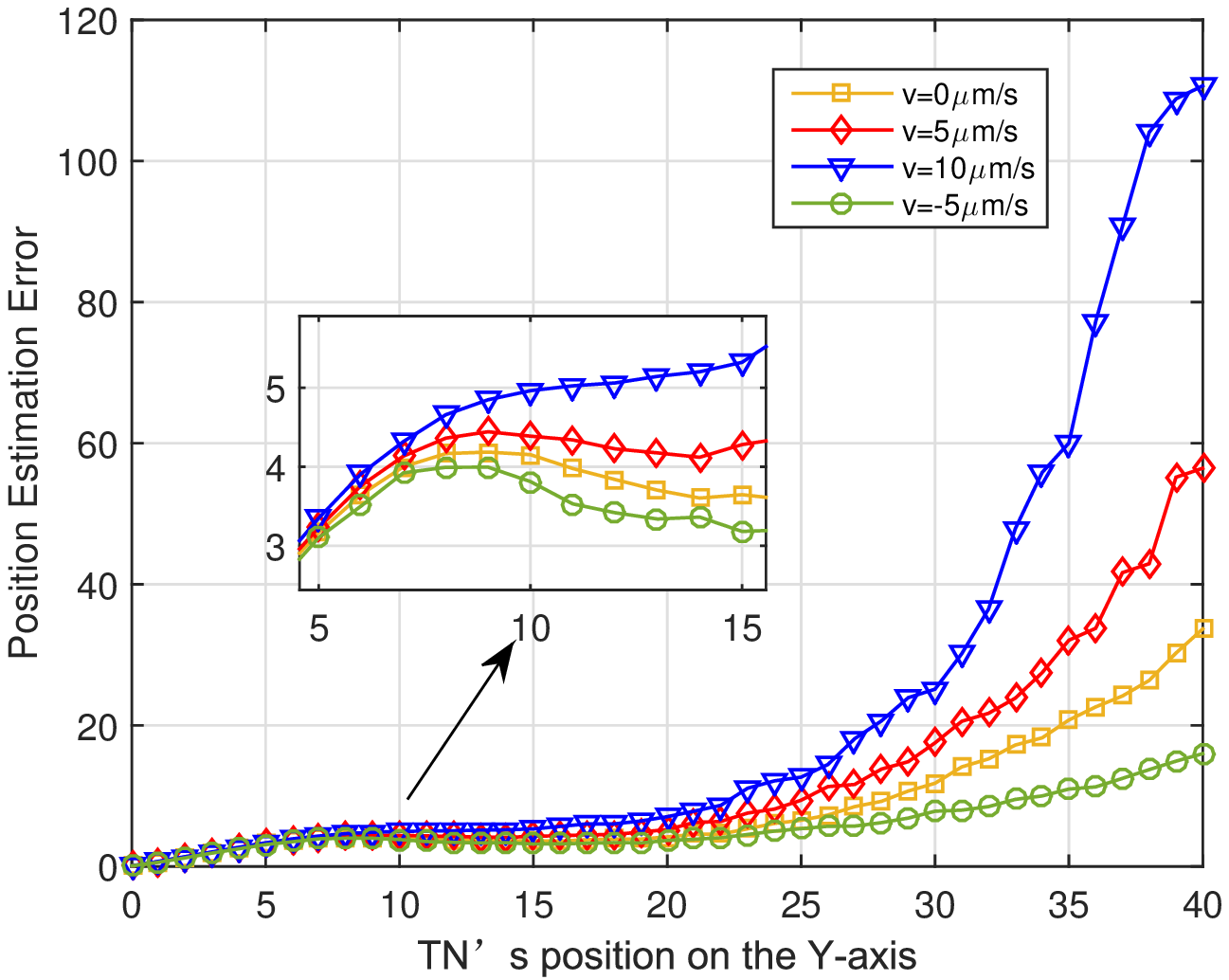}}	 	
	\caption{Location estimation error with respect to different impacting  factors.}	
	\label{fig:influence}
\end{figure*}

Factors such as RN's radius ($r_k$), the number of transmitted molecules ($Q$) and the diffusion coefficient ($D$) have influence on the number of received molecules. Therefore, the impact  of these parameters  on the accuracy of location estimation is evaluated which are shown in Fig. \ref{fig:influence}. The TN  is located on the Y axis ranging from (0,0,0) to (0,40,0).

 Fig. \ref{level.sub.1} shows the location estimation error with different RN radius, where    $Q=10000$ molecules, $D=100$ $\upmu$m$^2$/s and $r_k=1,2,3,4$ $\upmu$m. It is observed that the near-field and far-field results are different. In the far field (i.e., Y coordinate is greater than $20$), the location estimation error increases with respect to the TN-RN distance. The smaller RNs' radii become, the faster the estimation error increases. This is because as the distance from TN gets larger and the radius decreases, RNs with smaller radius receive less molecules, resulting in larger estimation error. But in the area near the receiver group, small radius leads to better performance. This is different from the single receiver scenarios. In the SIMO-MCvD systems, RNs ``compete'' for molecules, i.e., a molecule is absorbed by only one RN and then removed from the environment. When TN is located inbetween different RNs, a considerable part of the molecules are received by other RNs, resulting in inter-RN interference that degrades the estimation accuracy.

Fig. \ref{level.sub.2} shows the location estimation error with different transmitted molecules, where $r_k=1$ $\upmu$m, $D=100$ $\upmu$m$^2$/s and $Q=1000,5000,10000$ molecules. It is observed that as the actual distance increases or the number of released molecules decreases, the accuracy becomes worse. When the number of emitted molecules is higher than 5000, the estimation result is significantly better. Higher number of molecules brings significant advantages especially for the far-field.

Fig. \ref{level.sub.3} shows the location estimation error with different value of  diffusion coefficients, where $r_k=1$ $\upmu$m, $Q=10000$ molecules, and $D=100,200,300$ $\upmu$m$^2$/s. It is observed with larger value of $D$, the molecular motion is more active, making the number of received molecules increase, and the estimation result is better as a consequence.

The proposed CSL method is also suitable for scenarios with flow. Since the parameters are determined by curve fitting, the influence of flow can be adjusted with the auxiliary parameter $a_k$. Fig. \ref{level.sub.4} shows the location estimation error with different flow velocity, where $r_k=1$ $\upmu$m, $Q=10000$ molecules, $D=100$ $\upmu$m$^2$/s and ${\bf{v}}=(0,v,0), v\in\{-10,0,5,10\}   \upmu$m/s. It is clear that when the medium flows away from the receiver group, larger velocity brings negative effects due to the directional movement of molecules caused by the medium, and the number of received molecules is reduced. This phenomenon is more obvious in the far field, because the number of received molecules in the far field is too small, and the impact of small changes becomes significant.

\begin{figure}[htb]
  \centering
  \includegraphics[width=.8\columnwidth]{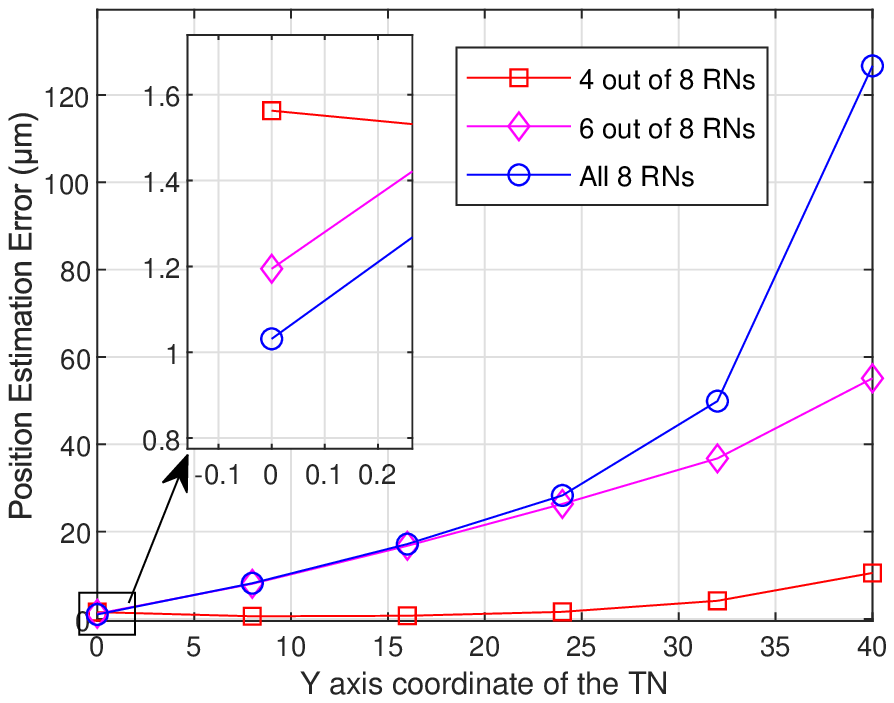}\\
  \caption{ Performance comparison for different selections out of 8 RNs.} \label{fig:cmp_best_8}
\end{figure}

Since different RNs may receive different numbers of molecules. Those RNs that are far away from the TN is dramatically affected by noise. Therefore, it is not always beneficial to use all the RNs in CSL. In Fig. 11, the RNs are located at the vertices of an origin-centred cube of side length 20 $\upmu$m. The location of the TN varies from (0,0,0) to (0,40,0) along the y-axis. It is shown that when 4 and 6 RNs with the largest number of received molecules are selected out of all the RNs in CSL, the performance is significantly improved. However,  when TN is located at the origin, using all 8 RNs shows the best accuracy.  This is because as TN's location changes, some RNs receive less molecules and the distance estimation error gets larger, due to increased distance from TN or blocking effect from other RNs. Therefore, avoiding using such RNs for CSL may improve the overall performance. This is significantly different from traditional wireless communications systems.

\subsection{Discussion on the Optimal RN Topology}

The problem of finding the optimal RN topology, i.e., the optimal ${\bf p}_1$ \ldots, ${\bf p}_K$, can be formulated as the following optimization problem

\begin{equation}\label{eqn:opt_pro_topo}
  \{ {\bf p}_1^*, \ldots, {\bf p}_K^* \} = \text {arg} \min \mathbb E \left \| \hat  {\bf p}_{\text T} - {\bf p}_{\text T}  \right \|_2,
\end{equation}
where $\hat  {\bf p}_{\text T}$ denotes the estimated TN location and the expectation is taken over $\bf p_{\text T}$.

It can be seen that the optimal solution to (\ref{eqn:opt_pro_topo}) is very difficult to obtain, since we generally cannot find an analytical  expression for $\mathbb E \left \| \hat  {\bf p}_{\text T} - \bf p_{\text T}  \right \|_2$.  In this work, we provide some conjectures on the optimal topology and more comprehensive discussions are deferred to future work.
\begin{conjecture} \label{conj:sym_topo}
Assuming the TN is uniformly distributed in the environment of interest, the solution to (\ref{eqn:opt_pro_topo}) has inherently symmetric structure.
\end{conjecture}

It is not easy to prove the conjecture rigorously, but some explanations are provided. Regarding Conjecture \ref{conj:sym_topo}, due to the symmetry of the 3D environment and uniformly distributed TN, a rotation of  the solution to (\ref{eqn:opt_pro_topo}) is also optimal. Therefore, the optimal solution should have inherent symmetry.

\begin{figure}[thb] 
	\centering  
	\subfigtopskip=2pt 
	\subfigcapskip=-5pt 
	\subfigure[]{		
		\includegraphics[width=.46\columnwidth]{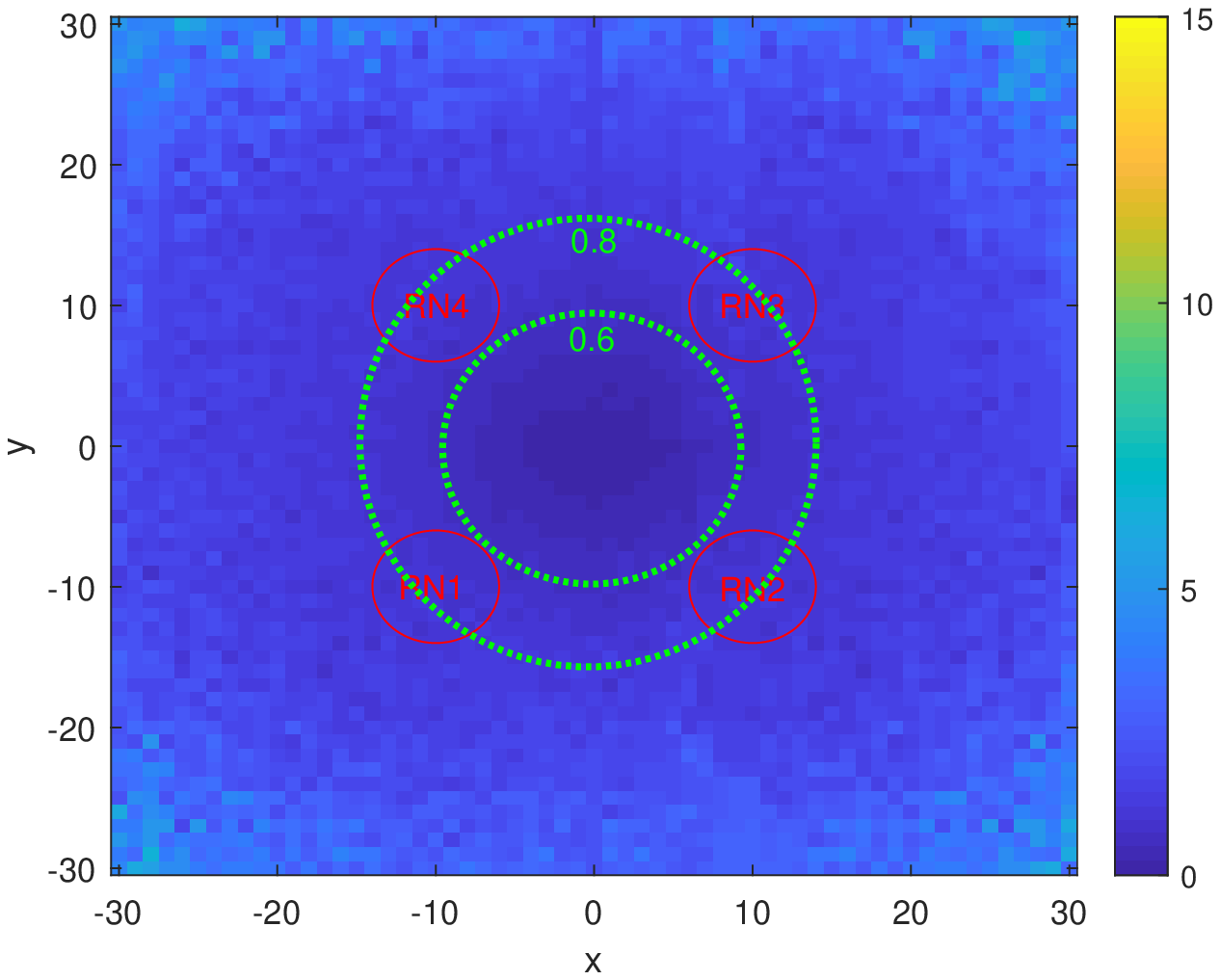}
		\label{fig:100_100_10}}	
	\subfigure[]{		
		\includegraphics[width=.46\columnwidth]{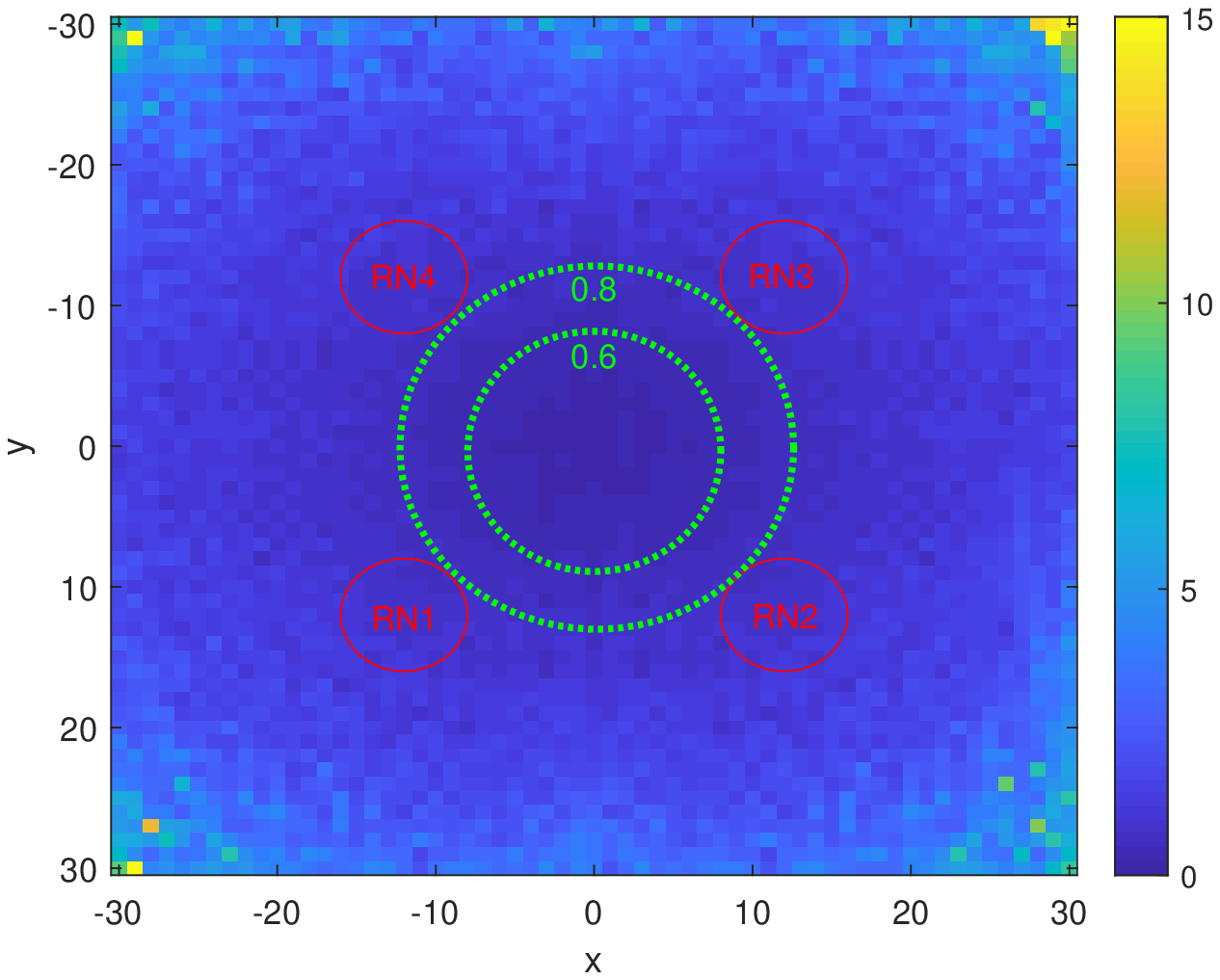}
	\label{fig:100_100_12}}	  	
	\caption{Location estimation error in the XoY plane, where the coordinates of the RNs are: (a) (-10,10,10), (10,10,-10), (10,-10,10), (-10,-10,-10) and (b) (-12,12,12), (12,12,-12), (12,-12,12), (-12,-12,-12)}	
	\label{fig:cmp_topo_10_12}	
\end{figure}

In the following,  we provide numerical results to show the complexity of the optimal topology problem.  In Fig. \ref{fig:100_100_10} and Fig. \ref{fig:100_100_12}, the locations of the four RNs are: (-10,10,10), (10,10,-10), (10,-10,10), (-10,-10,-10) and (-12,12,12), (12,12,-12), (12,-12,12), (-12,-12,-12), respectively, while the former configuration exhibits a better performance. The area where the estimation error is less than 0.6 $\upmu$m and 0.8 $\upmu$m are all enlarged compared with Fig. \ref{fig:100_100_5}, and the error level is also reduced in the middle area. This is because a distributed topology makes it easier for each RN to receive molecules emitted from a distant TN and  the inter-RN interference is weakened. However, when the distance between RNs continues to increase, as shown in Fig. \ref{fig:100_100_12}, the performance does not improve and it is worth noting that the error becomes large in the edge area, because some RNs are too far away from the TN to receive enough molecules when the TN is at the corner.


\begin{figure}[htb]
  \centering
  \includegraphics[width=.8\columnwidth]{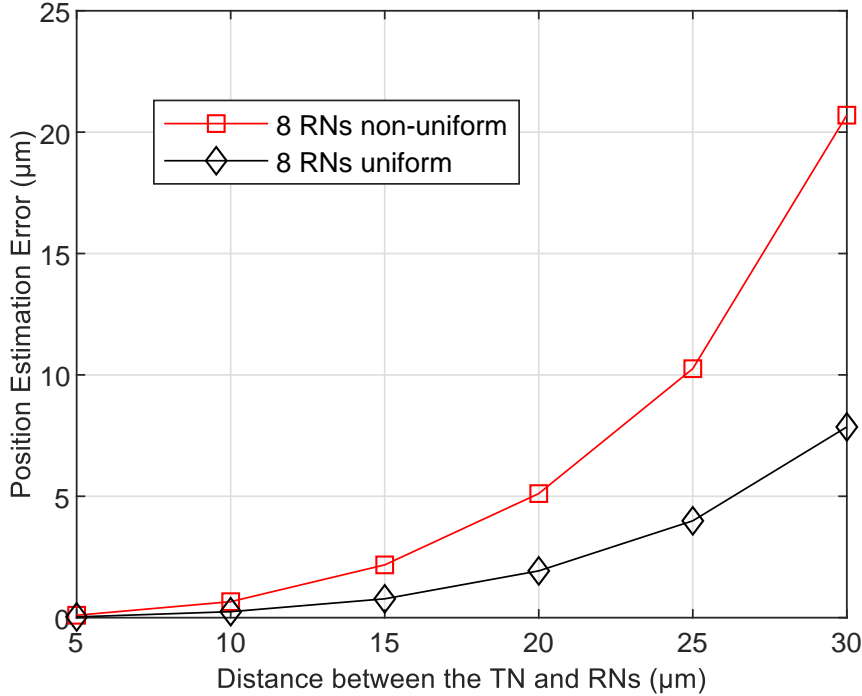}
  \caption{\color{blue} Location estimation error versus TN-RN distance where RNs are placed at a spherical surface centered at TN. The coordinates of RNs in uniform cases are $d*(-1,    -1,    -1)/\sqrt{3}$, $d*(-1,    -1    , 1)/\sqrt{3}$, $d*(-1 ,    1   , -1)/\sqrt{3}$, $d*(-1  ,   1   ,  1)/\sqrt{3}$, $d*(1    ,-1   , -1)/\sqrt{3}$, $d*(1   , -1    , 1)/\sqrt{3}$, $d*(1   ,  1   , -1)/\sqrt{3}$, $d*(1   ,  1  ,   1)/\sqrt{3}$ and in non-uniform cases are $d*(\frac{3}{5},    -\frac{3}{5},    -\frac{\sqrt{57}}{5})/\sqrt{3}$, $d*(\frac{3}{5},    -\frac{3}{5},    \frac{\sqrt{57}}{5})/\sqrt{3}$, $d*(\frac{3}{5},    \frac{3}{5},    \frac{\sqrt{57}}{5})/\sqrt{3}$, $d*(\frac{3}{5},    \frac{3}{5},    \frac{\sqrt{57}}{5})/\sqrt{3}$, $d*(1    ,-1   , -1)/\sqrt{3}$, $d*(1   , -1    , 1)/\sqrt{3}$, $d*(1   ,  1   , -1)/\sqrt{3}, d*(1   ,  1  ,   1)/\sqrt{3}$, where $d$ is the distance between the RNs and the TN.}
  \label{fig:dist_co_cmp_vs_d}
\end{figure}

Despite the overall evaluation in an area, now we focus on a specific TN.  In Fig. 13, the TN is fixed at the origin and all the RNs are deployed uniformly or non-uniformly at a spherical surface that is centered at the TN. As can be seen in the figure, the performance for 8 uniformly deployed RNs is significantly  better than its non-uniform counterpart, due to the reduced inter-RN interference. This indicates a distributed topology generally outperforms a co-located topology.

\section{Conclusion}
In this paper, we have presented a general model for MCvD systems with a single transmitter and multiple spherical absorption receivers. A novel distance estimation strategy is proposed and a cooperative source localization scheme is designed. Numerical results show the proposed scheme works well. Meanwhile, the performance of the proposed CSL is affected by RN's radius, the number of transmitted molecules and the diffusion coefficient. From the simulation results, the increase in the number of emitted molecules and the diffusion coefficient can significantly improve the accuracy of localization; in near-field communication, it suggests that RNs have small radius, as opposed to the far-field communication. This method is also suitable for flow scenarios and the directional movement of the fluid in the environment also improves the accuracy if the RNs are properly positioned. In addition, the complexity of this method can be reduced by finding the maximum sample interval. Future work could be done to improve the accuracy of the fitting equations to take the topology of multiple absorbing receivers into consideration.

\section*{Acknowledgement}
This work was supported in part by the National Nature Science Foundation of China (No. 61772243), Nature Science Foundation of Jiangsu Province (No. BK20170557), the Key Research \& Development Plan of Jiangsu Province (No. BE2018108), Six Talent Peak High Level Talent Plan Projects of Jiangsu Province (N0XYDXX-115) and Young Talent Project of Jiangsu University.

\bibliographystyle{IEEEtran}
\bibliography{miaoyuqi}

\end{document}